# Intelligent Detection and Mitigation of Carpet-Bombing DDoS Attacks in SDN Using Retrieval-Augmented Generation and Large Language Models

Mohammed N. Swileh (mohammedswileh2023@email.szu.edu.cn)[a,1], Shengli Zhang (zsl@szu.edu.cn)[a,*], Kai Lei (leik@pkusz.edu.cn)[b]

[a] *College of Electronics and Information Engineering, Shenzhen University, Shenzhen, 518060, China*

[b] *ICNLab, Shenzhen Graduate School, Peking University, Shenzhen, People's Republic of China*

[1] *First author*

[*] *Corresponding author*

**Abstract**

Software-Defined Networking (SDN) provides flexible and programmable network management; however, its centralized control architecture remains highly vulnerable to Distributed Denial-of-Service (DDoS) attacks, particularly Carpet-Bombing DDoS attacks that distribute malicious traffic across multiple targets to evade conventional detection mechanisms. In this paper, a Retrieval-Augmented Generation (RAG)-based framework is proposed for real-time detection and mitigation of Carpet-Bombing DDoS attacks in SDN environments. The proposed framework combines interface-level traffic features representation, semantic embedding generation, FAISS-based similarity retrieval, and Large Language Model (LLM)-driven contextual inference to classify traffic behavior without requiring conventional supervised model training or retraining. To evaluate the effectiveness of the proposed framework, extensive experiments were conducted under multiple Carpet-Bombing DDoS attack scenarios with different attack intensities. In addition, two traffic representation strategies, namely structured JSON-based representation and natural language-based representation (NLR), were investigated using multiple state-of-the-art LLMs. The experimental results demonstrate that the proposed framework achieved highly accurate and stable attack detection performance, while the framework configuration utilizing the Gemma-4-31B-IT model achieved the strongest overall detection results. Furthermore, real-time experiments confirmed the capability of the proposed framework to rapidly detect and mitigate Carpet-Bombing DDoS attacks while maintaining stable SDN network operation. The obtained results highlight the effectiveness of integrating RAG mechanisms with LLM for intelligent and adaptive SDN security analysis.

## 1. Introduction

In recent years, modern communication networks have experienced rapid growth driven by the increasing demand for cloud services, Internet of Things (IoT) applications, multimedia systems, and large-scale data-intensive environments. This rapid expansion has significantly increased network complexity, making traditional network architectures more difficult to manage, scale, and adapt to continuously changing traffic conditions [1, 2]. Conventional networking approaches typically rely on tightly coupled control and forwarding functionalities, which limits network flexibility and complicates network management and policy enforcement [3]. To address these limitations, SDN has emerged as a promising networking paradigm that separates the control plane from the data plane, enabling centralized and programmable network management [4, 5]. By introducing a logically centralized controller with global visibility of network behavior, SDN provides enhanced flexibility, simplified configuration, dynamic traffic engineering, and efficient resource management [6, 7]. Due to these advantages, SDN has been widely adopted in several modern networking environments, including cloud computing infrastructures, data centers, enterprise networks, and next-generation intelligent communication systems [8].

Despite the numerous advantages offered by SDN, its dynamic and programmable architecture also introduces several significant security challenges [9]. The centralized management model and continuous interaction between the control plane and data plane increase the attack surface of the network and create new opportunities for cyberattacks targeting different network components [10, 11]. Among these threats, DDoS attacks remain one of the most critical challenges in SDN environments due to their ability to rapidly consume network resources and disrupt normal network operations. In SDN-based infrastructures, DDoS attacks can affect multiple parts of the network simultaneously by overwhelming switches, saturating bandwidth links, exhausting flow table resources, and generating excessive communication with the controller [8, 12-14]. Such attacks may degrade network performance, increase packet processing delays, and interrupt legitimate services, potentially leading to partial or complete network failure.

However, the continuous evolution of DDoS attack strategies has introduced more sophisticated attack patterns that are increasingly difficult to detect using conventional defense mechanisms. Recently, several cybersecurity organizations, including Radware [15], Corero Network Security [16], A10 Networks [17], and NETSCOUT [18], have reported the emergence of Carpet-Bombing DDoS attacks as a serious threat to modern network infrastructures. Unlike traditional DDoS attacks that primarily focus on overwhelming a single victim or service, Carpet-Bombing DDoS attacks distribute malicious traffic across multiple destinations within the same network simultaneously using different traffic protocols and attack patterns [19-21]. In SDN environments, this attack behavior is particularly dangerous because each individual traffic flow may appear legitimate or low-volume when analyzed independently, making detection at the local level significantly more difficult. Nevertheless, when the distributed traffic is collectively analyzed, the attack generates substantial pressure on both the control plane and the data plane of the SDN architecture. Such attacks may trigger excessive flow requests, increase Packet-In messages between switches and the controller, consume bandwidth resources, overload forwarding devices, and exhaust computational resources such as CPU and memory. Consequently, Carpet-Bombing DDoS attacks can severely degrade network performance and potentially lead to large-scale service disruption or complete network failure.

Despite the growing body of research on DDoS detection in SDN environments, most existing studies primarily focus on conventional DDoS attacks that target a single victim or exhibit relatively distinguishable traffic behaviors [22-34]. Many proposed approaches rely on supervised machine learning, deep learning, or fine-tuning LLM techniques that require extensive training processes, large datasets, and periodic retraining to maintain detection performance under dynamic network conditions [21, 23-34]. In addition, many existing detection systems are designed to identify high-volume traffic anomalies or well-defined attack signatures, which may not be effective for distributed attack behaviors such as Carpet-Bombing DDoS attacks. However, in the case of Carpet-Bombing DDoS attacks, malicious traffic is distributed across multiple targets within the network, causing each individual traffic flow to appear relatively normal when analyzed separately [19-21]. As a result, traditional detection methods may struggle to accurately recognize the distributed nature of the attack or capture its aggregated impact on the overall SDN infrastructure. Furthermore, only a limited number of studies have specifically addressed the real-time detection and mitigation of Carpet-Bombing DDoS attacks in SDN environments, highlighting the need for more adaptive and intelligent detection frameworks capable of analyzing network-wide traffic behavior more effectively.

To address the limitations of existing approaches, this paper proposes a RAG-based framework for the real-time detection and mitigation of Carpet-Bombing DDoS attacks in SDN environments. The proposed framework integrates retrieval mechanisms with LLMs to analyze SDN interface traffic behavior and identify malicious activities without relying on conventional model training approaches. In addition, the framework utilizes structured traffic representations to improve contextual understanding and enhance detection performance in dynamic SDN environments.

The main contributions of this paper are summarized as follows:

1. A novel RAG-based framework is proposed for the real-time detection and mitigation of Carpet-Bombing DDoS attacks in SDN environments without relying on conventional supervised model training approaches.

2. The proposed framework leverages interface-level traffic features to identify Carpet-Bombing DDoS attack patterns across SDN infrastructures in real time, enabling early detection of malicious traffic behavior at the source interfaces.

3. The proposed framework investigates two traffic representation strategies, namely structured JSON-based representation and natural language-based representation, to analyze their impact on Carpet-Bombing DDoS attack detection performance in SDN environments.

4. Extensive experiments are conducted using multiple state-of-the-art LLMs under realistic Carpet-Bombing DDoS attack scenarios with different attack intensities in SDN environments.
5. Experimental results demonstrate that the proposed approach achieves highly accurate real-time detection performance while effectively mitigating the impact of Carpet-Bombing DDoS attacks on SDN infrastructures.

The remainder of this paper is organized as follows. Section 2 presents the related work on DDoS detection in SDN environments. Section 3 describes the proposed methodology and framework architecture. Section 4 discusses the experimental setup, performance evaluation, and obtained results. Finally, Section 5 concludes the paper and highlights potential future research directions.

## 2. Related Work

The growing impact of DDoS attacks in SDN environments has motivated extensive research efforts toward developing intelligent detection and mitigation mechanisms. Existing studies have explored a wide range of approaches, including traditional machine learning models, deep learning architectures, and more recently, LLM-based detection frameworks. These approaches differ in their feature extraction strategies, traffic representation methods, detection architectures, and mitigation capabilities. In addition, recent studies have begun investigating more sophisticated attack scenarios, such as distributed and Carpet-Bombing DDoS attacks, which introduce additional challenges for real-time detection in SDN infrastructures. Therefore, analyzing the existing literature is essential for understanding current research trends, identifying the limitations of existing approaches, and highlighting the remaining research challenges. Table 1 summarizes representative studies related to DDoS detection and mitigation in SDN environments.

Table 1: Summary of the Related Studies.

| Study | Details |
|---|---|
| [22] | Introduced a novel DDoS detection and mitigation framework designed for decentralized Software-Defined Networking (dSDN) environments. The proposed system utilizes lightweight port-level features and employs the pre-trained DeepSeek-v3 LLM for zero-training detection. By transforming port-level statistics into natural language prompts, the framework enables in-context learning to classify traffic as benign or malicious. The model achieved 99.99% accuracy, 99.97% precision, and 100% recall in detecting DDoS attacks. The system also integrates immediate mitigation by blocking malicious traffic at the source port, ensuring low-latency response and scalability. |
| [19] | Proposed a comprehensive architecture for malicious traffic detection powered by LLMs. They explored three roles that LLMs can play in traffic classification: Classifier, Encoder, and Predictor. The authors developed a framework that involves self-supervised pre-training, task-specific post-training, and detection processes. For the case study, they focused on detecting Carpet-Bombing DDoS attacks, utilizing the CIC-DDoS2019 and MAWI datasets for simulation, as well as a real-world NetFlow dataset from a nationwide operator. The system demonstrated a 35% improvement in detection accuracy compared to existing systems. |
| [23] | Proposed a BERT-based approach for detecting unseen attacks in SDN environments. The model transforms network flow data into a natural language processing format to leverage the capabilities of the pre-trained BERT-base-uncased model for attack detection. They utilized the InSDN dataset for training and testing. Random Forest was used for feature selection, ensuring efficient performance. The model demonstrated 99.96% accuracy, precision, recall, and F1-score, showcasing its ability to detect both known and unseen attacks. The approach also incorporated multi-flow analysis to improve the detection of complex and multi-stage attacks. |
| [24] | Proposed a lightweight and explainable LLM-based intrusion detection framework for SDN. They fine-tuned three LLMs—GPT Neo, Phi-2, and LLaMA2-7B—using 4-bit Quantized Low-Rank Adaptation (QLoRA) on the InSDN dataset. The approach involved transforming structured network features into natural language prompts for binary classification of normal and attack traffic. The results showed that LLaMA2-7B and Phi-2 performed exceptionally well, reaching 1.00 accuracy at the largest data scale of 350,000 samples. GPT Neo showed competitive performance at 350k but was slightly behind at smaller data sizes. |

[25] Proposed a method for detecting DDoS attacks in SDN environments using a two-stage detection approach. The first stage employs an anomaly detection method based on the Interquartile Range (IQR) to monitor the packet_in message rate from switches, with a dynamic threshold alarm algorithm to identify abnormal switches. The second stage involves a Deep Feature Fusion Convolutional Neural Network (DFFCNN) model for deep attack detection. The method was used on three datasets: CICIDS2017, CICIDS2018, and CICDDoS2019. Experimental results showed that the model achieved an average detection accuracy of 99.54%, with a false positive rate of 0.53%, offering an improvement of 1.65% in accuracy and a reduction of 1.38% in false positive rates compared to existing methods.

[20] Proposed DoLLM, a DDoS detection model leveraging LLMs for detecting Carpet-Bombing DDoS attacks. Unlike traditional detection systems that rely on explicit features, DoLLM reorganizes network flow data into Flow Sequences and uses LLMs' semantic understanding to enhance detection. The model applies a Flow Tokenizer to convert network flow data into token embeddings that LLMs can process, and then employs Bidirectional Self-Attention to analyze correlations between different flows. DoLLM was tested using the CICDDoS2019 dataset and real-world NetFlow data from a major ISP. Results showed that DoLLM outperformed other methods, with F1 scores improving by 33.3% in zero-shot scenarios and 20.6% in real ISP data.

[21] Introduced GMCB, a graph-based model designed for detecting Carpet-Bombing DDoS attacks. The model operates in three stages: graph construction, graph simplification, and anomaly detection. In the first stage, GMCB classifies traffic flows into long and short flows, then aggregates similar short flows to reduce graph complexity. In the second stage, the model simplifies the graph by applying connectivity analysis and DBSCAN clustering to remove unnecessary edges and vertices. In the final stage, GMCB identifies key vertices and performs anomaly detection by analyzing the interaction patterns of traffic flows. The approach was tested using a modified version of the CICDDoS2019 and MAWI datasets. The experimental results demonstrated that GMCB outperforms state-of-the-art models, achieving an AUC of 0.9894 and an F1 score of 0.9927 for Carpet-Bombing DDoS attack detection.

[26] Proposed an approach for detecting and mitigating DDoS attacks in SDN environments by combining Balanced Random Sampling (BRS) with Convolutional Neural Networks (CNN). The authors utilized the CICDDoS2019 dataset for their experiments. Their model demonstrated an accuracy of over 99.99% for binary classification and 98.64% for multi-class classification. Additionally, they incorporated mitigation strategies such as rate limiting and IP filtering to enhance network security.

[27] Introduced a Federated Learning (FedLAD) approach for detecting DDoS attacks in large-scale SDN. The model leverages decentralized machine learning, where each local controller in a multi-controller SDN architecture trains a local model using network traffic data without sharing sensitive data. The approach was evaluated using the CICDoS2017, CICDDoS2019, and InSDN datasets, achieving an accuracy of 98%. The system showed significant improvements in scalability and resource efficiency, with lower memory and CPU usage compared to other centralized approaches. FedLAD also demonstrated real-time DDoS detection with minimal resource consumption, making it suitable for large-scale SDN environments.

[28] Proposed a deep learning-based technique for detecting and mitigating DDoS attacks in SDN, focusing on both the control and data planes. The authors created their own dataset for training the model, which included traffic-based features such as unknown IP destination addresses and packet inter-arrival times. The model combined Autoencoder (AE) with Bidirectional Gated Recurrent Units (BGRU) and achieved an accuracy of 99.91% for the data plane and 99.89% for the control plane. Additionally, the approach incorporated mechanisms to block suspicious senders and mitigate attack effects.

[29] Introduced a deep neural network (DNN)-based approach for detecting and mitigating DDoS attacks in SDN environments. They used public datasets, including InSDN, CICIDS2018, and Kaggle DDoS, to train their model. The approach achieved detection accuracy rates of 99.98%, 100%, and 99.99%, respectively. The model demonstrated its effectiveness by identifying DDoS traffic with low loss rates. Additionally, the system leveraged the DNN model within an SDN controller to detect attacks in real-time, offering mitigation strategies to block malicious traffic.

[30] Proposed using LLMs like GPT-3.5, GPT-4, and Ada for detecting DDoS attacks in IoT systems. They employed both few-shot learning and fine-tuning methods for training their models. The datasets used in their

experiments were CICIDS2017 and Urban IoT. Their results showed that the LLMs achieved accuracies of about 95% for the CICIDS2017 dataset and nearly 96% for the Urban IoT dataset, surpassing traditional neural networks such as MLP. A key finding was that LLMs could provide reasoning behind their DDoS detection decisions, although fine-tuning occasionally led to hallucinations.

[31] Proposed a machine learning-based approach for detecting DDoS attacks targeting Software-Defined Vehicular Ad Hoc Networks (SD-VANETs). They used a custom-designed SD-VANET topology to create a dataset containing both normal traffic and DDoS attack traffic. The study applied the Minimum Redundancy Maximum Relevance (MRMR) feature selection algorithm to select the most relevant features from the dataset, and used Bayesian optimization for hyperparameter tuning of machine learning classifiers like KNN, SVM, and Decision Tree (DT). The results showed that the DT classifier with MRMR and Bayesian optimization achieved the highest accuracy of 99.35%.

[32] Proposed an efficient hybrid model for DDoS detection and classification in SDN-based IIoT networks. The model combines CNN and Long Short-Term Memory (LSTM) networks, with a feature selection process using the CICDDoS2019 dataset. The approach employs XGBoost for feature selection, which helps reduce the feature dimensions and improves the model's performance. By applying this hybrid CNN-LSTM model, the researchers achieved an accuracy of 99.50%, making it suitable for real-time applications.

[33] Proposed a DNN-based model for detecting and classifying DDoS attacks on network traffic. The model was trained using the CICDDoS2019 dataset, which includes both reflection-based and exploitation-based attack types. The DNN model achieved an accuracy of 99.99% for DDoS attack detection and 94.57% for classification of attack types. The study showed that the model performs well in detecting and categorizing DDoS attacks with high precision and recall.

[34] Proposed a machine learning-based approach for detecting and mitigating DDoS attacks in SDN. They created an SDN traffic dataset using the Mininet emulator, which includes 23 features such as packet count, byte count, and flow entries. The authors employed a hybrid model combining Support Vector Classifier (SVC) with Random Forest (RF) for traffic classification. Their model achieved an accuracy of 98.8%, with a low false alarm rate.

---

As summarized in Table 1, a significant portion of existing studies on DDoS detection in SDN environments primarily relies on traditional machine learning and deep learning approaches for identifying malicious traffic behavior [21, 25-29, 31-34]. Various machine learning and deep learning approaches have demonstrated high detection accuracy under different network conditions and attack scenarios. These approaches commonly utilize labeled datasets and engineered traffic features to classify network traffic as benign or malicious. However, despite their promising performance, most existing approaches depend heavily on supervised training processes, extensive datasets, and periodic retraining to maintain detection effectiveness under evolving network conditions. In addition, many traditional detection systems are designed to identify obvious traffic anomalies or predefined attack signatures, which may reduce their adaptability when dealing with highly distributed and dynamically evolving attack behaviors in SDN environments.

More recently, several studies have explored the use of LLMs for network intrusion and DDoS attack detection due to their strong contextual reasoning and pattern analysis capabilities [19, 20, 22-24, 30]. Existing LLM-based approaches have demonstrated promising detection performance by transforming network traffic features into natural language-based representations and utilizing prompt-based inference or fine-tuning strategies for traffic classification tasks. Compared with traditional machine learning techniques, LLM-based methods provide improved flexibility in analyzing complex traffic behaviors and unseen attack patterns. However, despite these advancements, many existing approaches still rely primarily on natural language traffic descriptions or computationally expensive fine-tuning procedures, which may limit scalability and adaptability in dynamic SDN environments. In addition, limited attention has been given to structured traffic representations and retrieval-driven architectures for SDN traffic analysis. Such approaches can enhance contextual reasoning by incorporating previously observed traffic behaviors during the inference process. Consequently, there remains a need for more adaptive LLM-based frameworks capable of supporting efficient and intelligent traffic analysis in SDN environments.

Although numerous studies have investigated DDoS detection and mitigation in SDN environments, most existing approaches primarily focus on conventional DDoS attacks that target a single victim or exhibit relatively distinguishable traffic behaviors [22-34]. In recent years, several studies have started exploring Carpet-Bombing DDoS attacks due to their highly distributed and stealthy characteristics [19-21]. These attacks distribute malicious traffic across multiple destinations within the same network, making each individual traffic flow appear relatively normal when analyzed independently. However, most existing Carpet-Bombing DDoS attacks detection studies have been evaluated using traditional network environments or general network flow datasets rather than SDN-specific infrastructures. Consequently, limited attention has been given to analyzing the impact of Carpet-Bombing DDoS attacks on SDN architectures, particularly regarding the interaction between the control plane and data plane, real-time traffic analysis, and source-level mitigation. Furthermore, the use of retrieval-driven LLM frameworks and structured traffic representations for detecting Carpet-Bombing DDoS attacks in SDN environments remains largely unexplored. Therefore, more adaptive and intelligent detection frameworks are still needed to effectively analyze and mitigate Carpet-Bombing DDoS attacks in real-time SDN infrastructures.

## 3. Methodology

This section describes the proposed methodology for real-time detection and mitigation of Carpet-Bombing DDoS attacks in SDN environments. The proposed framework combines traffic feature extraction, traffic representation, retrieval mechanisms, and LLMs to analyze SDN traffic behavior and identify malicious activities. In addition, the framework is designed to support real-time deployment within SDN infrastructures for efficient attack detection and mitigation.

### 3.1 Feature Extraction and Traffic Representation

In the proposed framework, interface-level traffic statistics are continuously collected from OpenFlow switches to monitor the behavior of each network interface individually within the SDN environment. The SDN controller periodically queries switch interfaces every 10 seconds and computes a set of traffic features separately for each interface to describe both incoming and outgoing network activities. The extracted features include Received_Packets, Received_Bytes, Received_Packets_Per_Second, Received_Bytes_Per_Second, Average_Received_Packet_Size, Sent_Packets, Sent_Bytes, Sent_Packets_Per_Second, Sent_Bytes_Per_Second, and Average_Sent_Packet_Size. These features were selected to capture traffic volume, packet transmission behavior, traffic rates, and packet size characteristics associated with both normal and malicious network activities. The 10-second monitoring interval was adopted to provide a balance between real-time responsiveness and traffic stability. Very short monitoring periods may produce unstable traffic observations and increase controller overhead due to frequent polling operations, whereas excessively long intervals may delay the detection of rapidly evolving attack activities.

Unlike conventional traffic analysis approaches that primarily focus on aggregated network flows, the proposed framework analyzes traffic behavior separately at each SDN switch's interface. This design enables the framework to monitor traffic distribution patterns across different interfaces in real time and identify abnormal traffic behavior directly at the source interfaces. Such interface-level visibility is particularly important for Carpet-Bombing DDoS attacks in SDN environments, where malicious traffic is distributed across multiple destinations and may appear relatively normal when observed as aggregated traffic flows. By analyzing traffic statistics independently for each switch's interface, the proposed framework can more effectively detect distributed attack activities at early stages and support mitigation before the attack traffic spreads across the SDN infrastructure.

#### 3.1.1 Network Traffic Representation

To enable effective traffic analysis within the proposed RAG-based framework, the extracted interface-level traffic statistics were transformed into structured traffic representations before being processed by the embedding and retrieval components. In this study, two different traffic representation strategies were investigated to analyze their effectiveness in representing SDN traffic behavior for Carpet-Bombing DDoS attack detection. The first approach utilizes a structured

JSON-based representation in which the extracted traffic features and their corresponding values are organized in a standardized format for each SDN interface. The second approach transforms the interface statistics into a natural language-based representation that describes the extracted traffic features and their associated statistical values for each SDN interface during the monitoring interval.

The structured JSON-based representation was designed to preserve the organization and relationships between traffic features while reducing semantic ambiguity during retrieval and inference operations. By maintaining a consistent feature structure, the JSON format facilitates more organized contextual processing and enables the retrieval mechanism and LLM to analyze traffic behavior more effectively across different SDN interfaces. In contrast, the natural language-based representation provides descriptive traffic context in a more human-readable format that can support semantic interpretation during LLM-based inference. Table 2 illustrates examples of the structured JSON-based representation and the natural language-based traffic representation utilized in the proposed framework.

Table 2: Examples of Network Traffic Representation Strategies

| Representation Type | | Example |
|---|---|---|
| **Structured JSON** | Retrieval data | {"interface_status": {"input_features": {"received": {"packets": 174, "bytes": 12776, "packets_per_second": 17.4, "bytes_per_second": 1277.6, "avg_packet_size": 73.4253}, "sent": {"packets": 193, "bytes": 15671, "packets_per_second": 19.3, "bytes_per_second": 1567.1, "avg_packet_size": 81.1969}}, "output_label": 0}} |
| | Test data | {"interface_status": {"input_features": {"received": {"packets": 955, "bytes": 1381981, "packets_per_second": 95.5, "bytes_per_second": 138198.1, "avg_packet_size": 1447.1005}, "sent": {"packets": 795, "bytes": 52672, "packets_per_second": 79.5, "bytes_per_second": 5267.2, "avg_packet_size": 66.2541}}}} |
| **Natural Language** | Retrieval data | The interface received 174 packets totaling 12776 bytes with a rate of 17.4 packets per second and 1277.6 bytes per second. The average received packet size was 73.4253 bytes. It transmitted 193 packets totaling 15671 bytes at a rate of 19.3 packets per second and 1567.1 bytes per second with an average transmitted packet size of 81.1969 bytes. The interface label is 0. |
| | Test data | The interface received 955 packets totaling 1381981 bytes with a rate of 95.5 packets per second and 138198.1 bytes per second. The average received packet size was 1447.1005 bytes. It transmitted 795 packets totaling 52672 bytes at a rate of 79.5 packets per second and 5267.2 bytes per second with an average transmitted packet size of 66.2541 bytes. |

### 3.2 Proposed RAG-Based Detection Framework

The proposed framework utilizes a RAG-based architecture to analyze SDN interface traffic behavior and detect Carpet-Bombing DDoS attacks in real time without relying on conventional supervised model training. The framework combines structured traffic representations, semantic embedding generation, similarity-based retrieval, and LLM inference to classify network traffic as benign or malicious. Initially, the traffic dataset is divided into retrieval and testing subsets, after which embedding vectors are generated and indexed using the FAISS similarity search framework. During the inference stage, each test query is transformed into an embedding vector and compared against previously stored traffic representations to retrieve the most semantically similar examples from both benign and attack classes. Subsequently, the retrieved contextual samples and the target query are combined into a structured prompt and provided to the LLM for final traffic classification. Figure 1 illustrates the overall architecture and operational workflow of the proposed RAG-based detection framework.

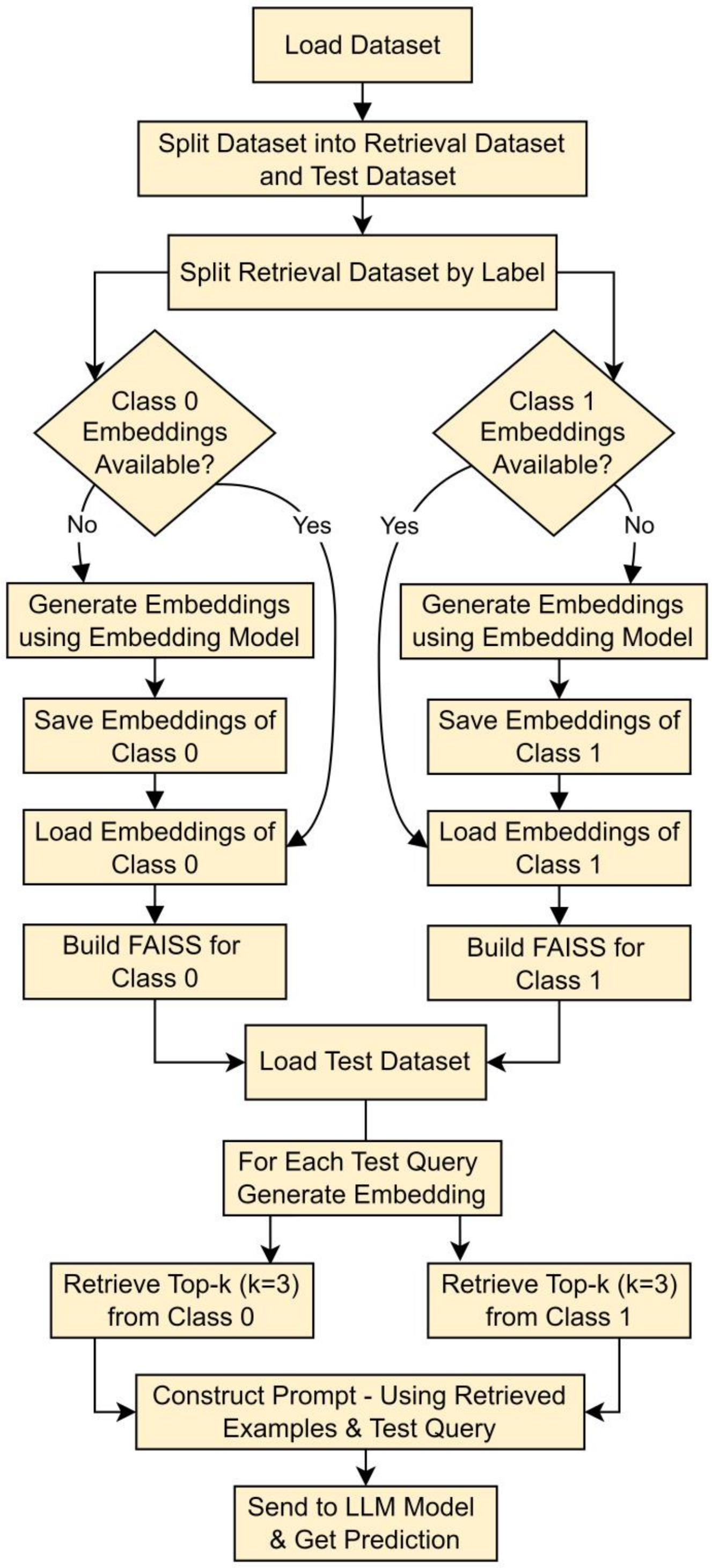

Figure 1: Operational Workflow of the Proposed RAG-Based Detection Framework

### 3.2.1 Retrieval Dataset Organization

To support the retrieval process within the proposed RAG-based framework, the generated dataset was divided into two main subsets: an 80% retrieval dataset and a 20% testing dataset. The retrieval dataset was utilized as the contextual knowledge source for similarity search operations, whereas the testing dataset was used to evaluate the detection performance of the proposed framework. Each traffic sample within the dataset represents the statistical behavior of an individual SDN switch interface during a specific monitoring interval and contains the extracted interface-level traffic features together with the corresponding output label indicating either benign traffic or Carpet-Bombing DDoS attack traffic.

After organizing the retrieval dataset, the traffic samples were further organized according to their class labels into two separate groups representing benign traffic samples and attack traffic samples. This separation enables the framework to independently retrieve semantically similar examples from both classes during the inference stage, allowing the LLM to analyze contextual similarities and differences between normal and malicious traffic behaviors more effectively. By organizing the retrieval dataset into class-specific traffic representations, the proposed framework improves retrieval efficiency and supports more structured contextual analysis during the attack detection process.

#### 3.2.2 Embedding Generation and FAISS Indexing

After constructing the retrieval dataset, each traffic representation was transformed into a dense embedding vector to enable semantic similarity analysis during the retrieval process. In the proposed framework, the paraphrase-MiniLM-L6-v2 embedding model was utilized to generate vector representations for both benign and attack traffic samples. The embedding process converts each traffic representation into a high-dimensional numerical vector that captures the semantic relationships and contextual characteristics of the observed SDN traffic behavior. This transformation enables the framework to compare traffic samples based on semantic similarity rather than relying solely on exact feature matching.

To support efficient similarity search operations, the generated embedding vectors were indexed using the Facebook AI Similarity Search (FAISS) framework. Separate FAISS indices were constructed for benign traffic samples and attack traffic samples to facilitate independent retrieval from both traffic classes during the inference stage. Organizing the embeddings into class-specific indices allows the proposed framework to efficiently retrieve the most semantically similar traffic examples associated with both normal and malicious network behavior. During the retrieval process, the indexed embedding vectors enable fast nearest-neighbor similarity search operations, which are essential for supporting real-time traffic analysis and attack detection in SDN environments.

#### 3.2.3 Query Embedding and Similarity Retrieval

During the inference stage, each incoming SDN interface traffic representation from the testing dataset is treated as a target query for the retrieval process. Initially, the query representation is transformed into an embedding vector using the same embedding model utilized during the retrieval dataset construction phase to ensure embedding consistency between stored and incoming traffic samples. The generated query embedding captures the semantic characteristics and statistical behavior of the observed interface traffic, enabling similarity-based comparison against previously indexed traffic representations.

After generating the query embedding, similarity search operations are performed using the FAISS indexing framework to retrieve the most semantically similar traffic samples from both the benign and attack retrieval indices. In this study, the retrieval process returns the top three most semantically similar benign traffic samples and the top three most semantically similar attack traffic samples for each target query. The retrieval process identifies the nearest neighboring traffic representations based on vector similarity distances, allowing the framework to obtain contextual examples associated with traffic behaviors that closely resemble the target query. The retrieved traffic samples provide contextual knowledge regarding previously observed normal and malicious traffic patterns, which assist the LLM during the inference process in distinguishing between benign SDN traffic behavior and potential Carpet-Bombing DDoS attack activities.

#### 3.2.4 Prompt Construction and LLM Inference

After retrieving the most semantically similar traffic samples, the retrieved benign and attack examples are combined with the target SDN interface traffic representation to construct the final prompt provided to the LLM, as shown in Figure 2. The prompt contains contextual traffic examples retrieved from both traffic classes together with the target query, enabling the LLM to analyze semantic similarities and behavioral differences between normal and malicious SDN traffic patterns.

Task: Detect whether the interface status observed during the last ten seconds indicates an attack or normal behavior. Analyze the provided examples labeled interface status, then classify the target interface status accordingly.

Labeled interface status:

{"interface_status": {"input_features": {"received": {"packets": 4932, "bytes": 7084704, "packets_per_second": 493.2, "bytes_per_second": 708470.4, "avg_packet_size": 1436.4769}, "sent": {"packets": 1, "bytes": 98, "packets_per_second": 0.1, "bytes_per_second": 9.8, "avg_packet_size": 98.0}}, "output_label": 0}}

{"interface_status": {"input_features": {"received": {"packets": 7090, "bytes": 10173308, "packets_per_second": 709.0, "bytes_per_second": 1017330.8, "avg_packet_size": 1434.8812}, "sent": {"packets": 8, "bytes": 432, "packets_per_second": 0.8, "bytes_per_second": 43.2, "avg_packet_size": 54.0}}, "output_label": 0}}

{"interface_status": {"input_features": {"received": {"packets": 3837, "bytes": 5493306, "packets_per_second": 383.7, "bytes_per_second": 549330.6, "avg_packet_size": 1431.6669}, "sent": {"packets": 1, "bytes": 98, "packets_per_second": 0.1, "bytes_per_second": 9.8, "avg_packet_size": 98.0}}, "output_label": 0}}

{"interface_status": {"input_features": {"received": {"packets": 209303, "bytes": 34909398, "packets_per_second": 20930.3, "bytes_per_second": 3490939.8, "avg_packet_size": 166.7888}, "sent": {"packets": 3, "bytes": 126, "packets_per_second": 0.3, "bytes_per_second": 12.6, "avg_packet_size": 42.0}}, "output_label": 1}}

{"interface_status": {"input_features": {"received": {"packets": 279983, "bytes": 46206566, "packets_per_second": 27998.3, "bytes_per_second": 4620656.6, "avg_packet_size": 165.0335}, "sent": {"packets": 1, "bytes": 42, "packets_per_second": 0.1, "bytes_per_second": 4.2, "avg_packet_size": 42.0}}, "output_label": 1}}

{"interface_status": {"input_features": {"received": {"packets": 334864, "bytes": 56956136, "packets_per_second": 33486.4, "bytes_per_second": 5695613.6, "avg_packet_size": 170.0874}, "sent": {"packets": 2, "bytes": 84, "packets_per_second": 0.2, "bytes_per_second": 8.4, "avg_packet_size": 42.0}}, "output_label": 1}}

Target interface status:

{"interface_status": {"input_features": {"received": {"packets": 4386, "bytes": 6317894, "packets_per_second": 438.6, "bytes_per_second": 631789.4, "avg_packet_size": 1440.4683}, "sent": {"packets": 2, "bytes": 148, "packets_per_second": 0.2, "bytes_per_second": 14.8, "avg_packet_size": 74.0}}}}

Instructions: Only answer with one number, the label of the target interface status: 0 for Benign, 1 for Attack. Do not explain.

Task: Detect whether the interface status observed during the last ten seconds indicates an attack or normal behavior. Analyze the provided examples labeled interface status, then classify the target interface status accordingly.

Labeled interface status:

The interface received 4360 packets totaling 6248696 bytes with a rate of 436.0 packets per second and 624869.6 bytes per second. The average received packet size was 1433.1872 bytes. It transmitted 6 packets totaling 444 bytes at a rate of 0.6 packets per second and 44.4 bytes per second with an average transmitted packet size of 74.0 bytes. The interface label is 0.

The interface received 4384 packets totaling 6288876 bytes with a rate of 398.5455 packets per second and 571716.0 bytes per second. The average received packet size was 1434.5064 bytes. It transmitted 1 packets totaling 74 bytes at a rate of 0.0909 packets per second and 6.7273 bytes per second with an average transmitted packet size of 74.0 bytes. The interface label is 0.

The interface received 4378 packets totaling 6310388 bytes with a rate of 437.8 packets per second and 631038.8 bytes per second. The average received packet size was 1441.386 bytes. It transmitted 2 packets totaling 196 bytes at a rate of 0.2 packets per second and 19.6 bytes per second with an average transmitted packet size of 98.0 bytes. The interface label is 0.

The interface received 326982 packets totaling 54194816 bytes with a rate of 32698.2 packets per second and 5419481.6 bytes per second. The average received packet size was 165.7425 bytes. It transmitted 9 packets totaling 378 bytes at a rate of 0.9 packets per second and 37.8 bytes per second with an average transmitted packet size of 42.0 bytes. The interface label is 1.

The interface received 263418 packets totaling 44260020 bytes with a rate of 26341.8 packets per second and 4426002.0 bytes per second. The average received packet size was 168.022 bytes. It transmitted 4 packets totaling 168 bytes at a rate of 0.4 packets per second and 16.8 bytes per second with an average transmitted packet size of 42.0 bytes. The interface label is 1.

The interface received 327630 packets totaling 55282084 bytes with a rate of 32763.0 packets per second and 5528208.4 bytes per second. The average received packet size was 168.7333 bytes. It transmitted 27 packets totaling 1134 bytes at a rate of 2.7 packets per second and 113.4 bytes per second with an average transmitted packet size of 42.0 bytes. The interface label is 1.

Target interface status:

The interface received 4386 packets totaling 6317894 bytes with a rate of 438.6 packets per second and 631789.4 bytes per second. The average received packet size was 1440.4683 bytes. It transmitted 2 packets totaling 148 bytes at a rate of 0.2 packets per second and 14.8 bytes per second with an average transmitted packet size of 74.0 bytes.

Instructions: Only answer with one number, the label of the target interface status: 0 for Benign, 1 for Attack. Do not explain.

Figure 2: Examples of structured JSON-based and natural language-based traffic representation prompts utilized in the proposed framework.

As illustrated in Figure 2, the prompt structure was designed to provide clear contextual organization between benign traffic samples, attack traffic samples, and the target query representation. In addition, explicit task instructions were incorporated into the prompt to guide the LLM toward binary traffic classification by determining whether the observed traffic behavior corresponds to normal SDN activity or a potential Carpet-Bombing DDoS attack. This structured contextual organization assists the LLM in performing more consistent traffic analysis during inference operations.

In this study, multiple state-of-the-art LLMs were evaluated to investigate their effectiveness in SDN traffic analysis tasks, while Gemma-4-31B-IT was adopted as the primary model within the proposed framework due to its strong inference performance during experimental evaluation. During inference, the LLM analyzes the constructed prompt and generates the final prediction label associated with the target traffic representation. By integrating semantic retrieval with LLM-based inference, the proposed framework enhances contextual traffic understanding and improves the detection of Carpet-Bombing DDoS attack behavior in SDN environments.

**3.3 Real-time Proposed RAG System to detect and mitigate Carpet-Bombing DDoS attack**

To enable real-time detection and mitigation of Carpet-Bombing DDoS attacks in SDN environments, the proposed RAG-based framework was integrated directly within the SDN controller to continuously monitor interface-level traffic behavior and analyze network activities during operation. Figure 3 illustrates the real-time operational workflow of the proposed framework. During system initialization, the SDN controller loads the retrieval dataset, embedding model, precomputed embedding vectors, FAISS similarity indices, and the LLM inference components required for traffic analysis. After initialization, the framework starts a continuous monitoring process in which the controller periodically requests interface statistics from OpenFlow switches every 10 seconds to observe the behavior of each SDN interface individually.

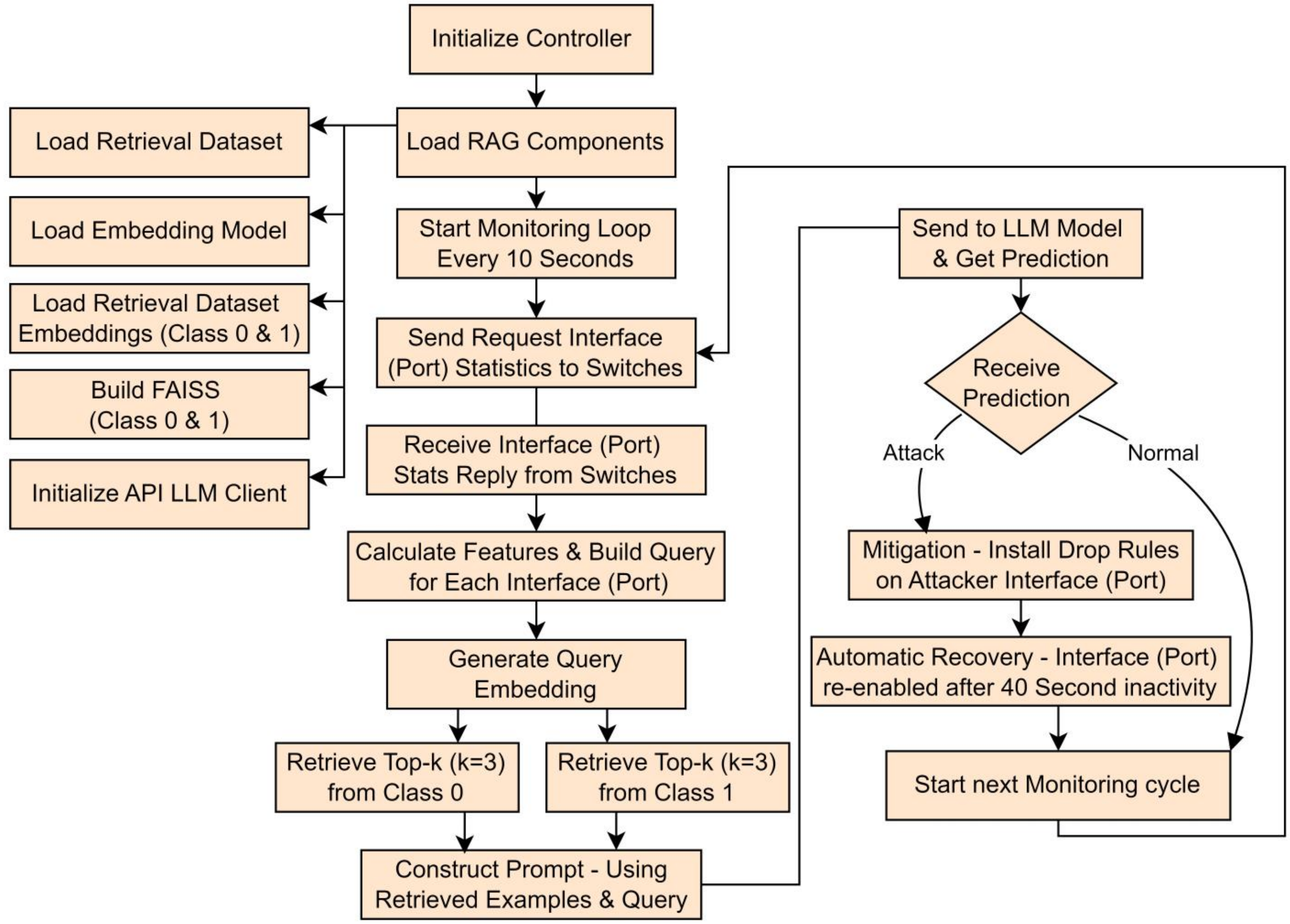

Figure 3: Real-Time Operational Workflow of the Proposed RAG-Based Detection and Mitigation Framework

After receiving the interface statistics from the switches, the controller computes the extracted traffic features for each interface and transforms them into the selected traffic representation format utilized by the proposed framework. Subsequently, the generated traffic representation is converted into an embedding vector using the embedding model and compared against the indexed retrieval dataset through the FAISS similarity search mechanism. The retrieval stage returns the top three most semantically similar benign traffic samples together with the top three most semantically similar attack traffic samples associated with the target interface behavior. The retrieved contextual examples and the target query are then combined into a structured prompt and provided to the LLM to classify the observed traffic behavior as either benign traffic or a Carpet-Bombing DDoS attack.

Once the prediction result is generated, the SDN controller determines the appropriate network response according to the inferred traffic label. If the traffic is classified as benign, the monitoring process continues normally for the subsequent monitoring interval. However, if the traffic is identified as a Carpet-Bombing DDoS attack, mitigation actions are immediately applied at the corresponding source interface by installing OpenFlow drop rules to block malicious traffic transmission. Applying mitigation directly at the source interface enables the proposed framework to reduce the spread of attack traffic across the SDN infrastructure and minimize the impact of distributed attack behavior on both the control plane and the data plane.

To avoid permanent interface blocking and maintain normal network operation, the proposed framework also incorporates an automatic recovery mechanism in which blocked interfaces are re-enabled after a predefined inactivity period of 40 seconds. Following the mitigation or recovery process, the framework proceeds to the next monitoring cycle to support continuous real-time traffic analysis, attack detection, and adaptive mitigation within the SDN environment.

## 4. Results and Discussion

This section presents the experimental evaluation and performance analysis of the proposed RAG-based framework for detecting and mitigating Carpet-Bombing DDoS attacks in SDN environments. Initially, the experimental setup, dataset construction process, and evaluation metrics utilized in this study are described. Subsequently, the performance results of the proposed framework are discussed and analyzed under different traffic representation strategies. In addition, real-time experiments are conducted to evaluate the effectiveness of the proposed framework in detecting and mitigating Carpet-Bombing DDoS attacks within the SDN environment under different attack intensities. Finally, the detection performance of the proposed framework is comparatively analyzed using different traffic representation strategies and multiple LLMs.

### 4.1 Experimental Setup

The experimental environment utilized in this study was designed to support both SDN traffic dataset generation and evaluation of the proposed RAG-based detection and mitigation framework. The SDN topology and network emulation environment were implemented using Mininet, while the Ryu controller was employed to manage OpenFlow switches and perform traffic monitoring operations. Normal network traffic was generated using the iPerf tool, whereas Carpet-Bombing DDoS attack traffic was generated using hping3 under different attack intensity scenarios. In addition, Jupyter Notebook was utilized to perform offline experiments and evaluate the proposed framework using the constructed traffic dataset. Two separate high-performance servers were utilized throughout this study. The first server was dedicated to dataset generation and traffic collection processes. This server operated using Ubuntu 20.04.6 LTS with Linux Kernel 5.15.0-139-generic and was equipped with dual Intel Xeon Gold 6230R processors providing 52 cores and 104 threads at 2.10 GHz, together with approximately 1.2 TiB of RAM. The second server was utilized to implement and evaluate the complete proposed RAG-based framework, including embedding generation, FAISS indexing, LLM inference, and real-time detection and mitigation operations. This server operated using Ubuntu 22.04 with Linux Kernel 6.5.0-44-generic and was equipped with an AMD EPYC 9654 processor providing 96 cores and 192 threads with frequencies ranging from 1.50 GHz to 3.70 GHz, 503 GiB of RAM, and an NVIDIA A100 80 GB PCIe GPU utilizing CUDA 12.4 and NVIDIA Driver version 550.90.07.

### 4.2 Dataset Construction

To construct the SDN traffic dataset utilized in this study, a custom SDN topology was designed and implemented using the Mininet network emulator, as illustrated in Figure 4. The topology consisted of a single SDN controller connected to six OpenFlow switches, where each switch was further connected to five hosts. Several hosts were configured as dedicated service servers to generate different types of legitimate network traffic. Specifically, hosts h5, h15, and h25 were configured as TCP servers, hosts h10, h20, and h30 operated as UDP servers, while hosts h9 and h24 functioned as HTTP servers. The remaining hosts operated as client devices that continuously generated normal background traffic toward the available servers throughout the experiments. To simulate realistic communication behavior within the SDN environment, the client hosts generated multiple types of normal network traffic, including ICMP ping requests, TCP connections, UDP flows, and HTTP requests. Both the communicating client hosts and the target servers were selected dynamically during the experiments to increase traffic diversity and reduce repetitive communication patterns. This experimental setup allowed the generated dataset to represent a wide range of legitimate network activities and traffic behaviors, reflecting the natural diversity of real-world network environments.

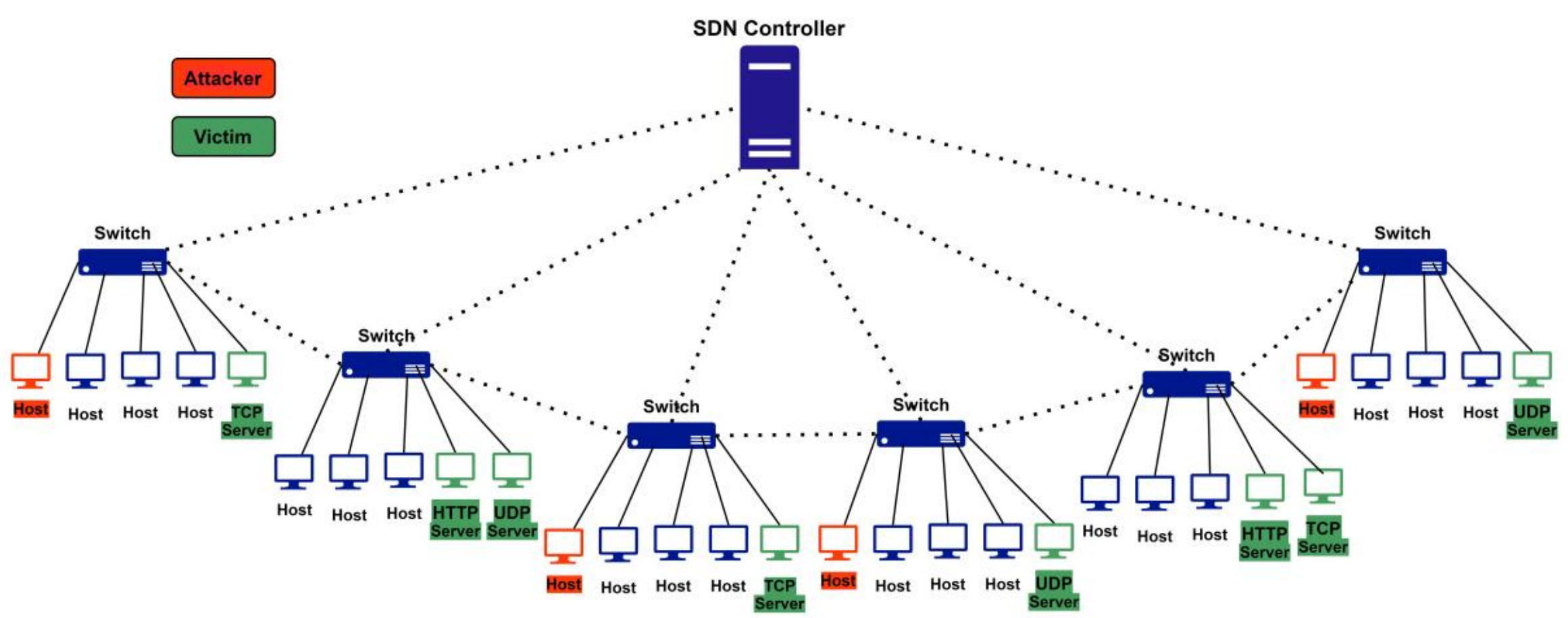

Figure 4: SDN experimental topology utilized in this study

To evaluate the proposed framework under Carpet-Bombing DDoS attack conditions, multiple attacker hosts were deployed within the SDN topology, as illustrated in Figure 4. Four hosts were configured as attackers that simultaneously generated distributed attack traffic toward multiple service servers using different attack protocols. Each attacker host targeted two different servers concurrently using separate attack types. For example, an attacker host could simultaneously launch a TCP attack toward a TCP server while generating a UDP attack toward a UDP server during the same attack interval. This distributed attack behavior was designed to emulate realistic Carpet-Bombing DDoS attack patterns in which malicious traffic is dispersed across multiple network destinations instead of concentrating on a single victim. To analyze the effectiveness of the proposed framework under different attack intensities, four independent Carpet-Bombing DDoS attack scenarios were conducted using different traffic generation rates. The first scenario generated attack traffic at approximately 100,000 packets per second, while the second, third, and fourth scenarios generated approximately 66,666, 40,000, and 20,000 packets per second, respectively. During all attack scenarios, the generated traffic samples were systematically labeled according to the known attack timing and source interfaces to ensure accurate differentiation between benign and malicious traffic behaviors within the constructed dataset.

The dataset collection process was conducted continuously over approximately 72 hours under both normal traffic conditions and Carpet-Bombing DDoS attack scenarios. During this process, interface-level traffic statistics were aggregated every 10 seconds from the SDN switch interfaces. The constructed dataset contains ten interface-level traffic features discussed in Section 3.1, which were designed to capture network traffic behavior associated with both benign and malicious activities within the SDN environment. In total, 100,000 traffic records were collected, where 55,118 samples represent normal network traffic and 44,882 samples correspond to Carpet-Bombing DDoS attack activities, as illustrated in Figure 5. Furthermore, the attack samples were collected from the four attack scenarios using approximately balanced distributions to ensure fair evaluation of the proposed framework under different Carpet-Bombing DDoS attack intensities.

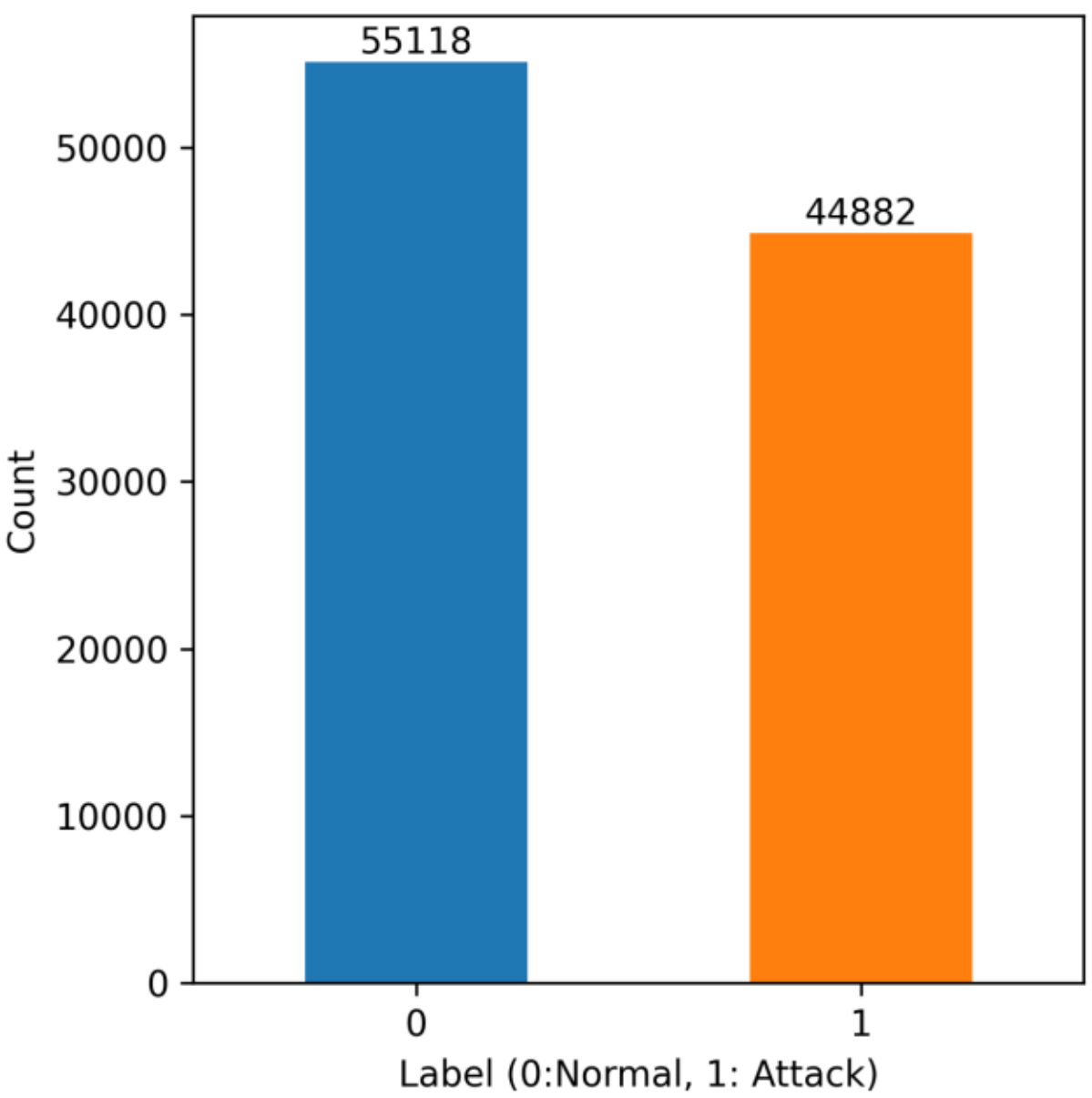

Figure 5: Distribution of benign and Carpet-Bombing DDoS attack samples in the dataset.

### 4.3 Performance Evaluation Metrics

To evaluate the performance of the proposed RAG-based framework for detecting Carpet-Bombing DDoS attacks in SDN environments, several commonly utilized classification metrics were employed in this study, including Accuracy, Precision, Recall, F1-Score, Confusion Matrix, and Receiver Operating Characteristic (ROC) with Area Under the Curve (AUC). In the following equations, TP, TN, FP, and FN denote true positives, true negatives, false positives, and false negatives, respectively.

- Accuracy (ACC):

$$Accuracy = \frac{TP + TN}{TP + TN + FP + FN} \quad (1)$$

- Precision (PRE):

$$Precision = \frac{TP}{TP + FP} \quad (2)$$

- Recall (REC):

$$Recall = \frac{TP}{TP + FN} \quad (3)$$

- F1-Score (F1):

$$F1 - Score = 2 * \frac{Precision * Recall}{Precision + Recall} \quad (4)$$

In addition, Confusion Matrix and ROC-AUC analyses were utilized to further evaluate the classification performance and discrimination capability of the proposed framework.

### 4.4 Performance Evaluation of the Proposed RAG-Based Framework

The performance of the proposed RAG-based framework was evaluated using the constructed traffic dataset under both normal traffic conditions and Carpet-Bombing DDoS attack scenarios. The framework utilizes the Gemma-4-31B-IT model as the core inference component for contextual traffic analysis and attack classification using the two proposed traffic representation strategies, namely the structured JSON-based representation and the natural language-based representation. Several performance evaluation metrics, including Accuracy, Precision, Recall, F1-Score, Confusion Matrix, ROC curve, and AUC, were utilized to assess the effectiveness of the proposed framework in distinguishing between benign and malicious SDN traffic behaviors. The obtained results demonstrate that the proposed framework achieved high detection performance using both traffic representation strategies, indicating the effectiveness of the proposed RAG-based approach for Carpet-Bombing DDoS attack detection in SDN environments. Table 3 summarizes the overall performance results obtained during the evaluation process.

Table 3: Performance results of the proposed RAG-based framework under different traffic representation strategies.

| Representation | Model | Accuracy | Precision | Recall | F1-score | Average Request Time |
|---|---|---|---|---|---|---|
| JSON | Gemma-4-31B-IT | 99.94% | 99.92% | 99.93% | 99.93% | 2.10/s |
| NLR | Gemma-4-31B-IT | 99.93% | 99.86% | 99.98% | 99.92% | 2.08/s |

Table 3 demonstrates that the proposed RAG-based framework achieved excellent and highly reliable detection performance under both traffic representation strategies. Using the structured JSON-based representation, the framework achieved highly accurate traffic classification performance with an accuracy of 99.94%, precision of 99.92%, recall of 99.93%, and F1-score of 99.93%, while maintaining an average request processing time of approximately 2 seconds. Similarly, the natural language-based representation also achieved strong detection capability with an accuracy of 99.93%, precision of 99.86%, recall of 99.98%, and F1-score of 99.92%, with an average request processing time of approximately 2 seconds. The obtained confusion matrices further demonstrate the effectiveness of the proposed framework in distinguishing between benign and malicious SDN traffic behaviors with very limited misclassification rates under both traffic representation strategies. Specifically, the structured JSON-based representation achieved 11,017 true positives and 8,970 true negatives with only 7 false positives and 6 false negatives, indicating highly stable classification behavior during inference operations. Likewise, the natural language-based representation achieved 11,011 true positives and 8,974 true negatives with only 13 false positives and 2 false negatives, further confirming the robustness of the proposed framework under different traffic representation formats. Furthermore, the ROC curve analysis achieved an AUC value of 1.00 for both traffic representation strategies, demonstrating the strong discrimination capability of the proposed framework in separating benign SDN traffic from Carpet-Bombing DDoS attack activities. Figures 6 and 7 illustrate the obtained confusion matrices and ROC curve results for both traffic representation strategies.

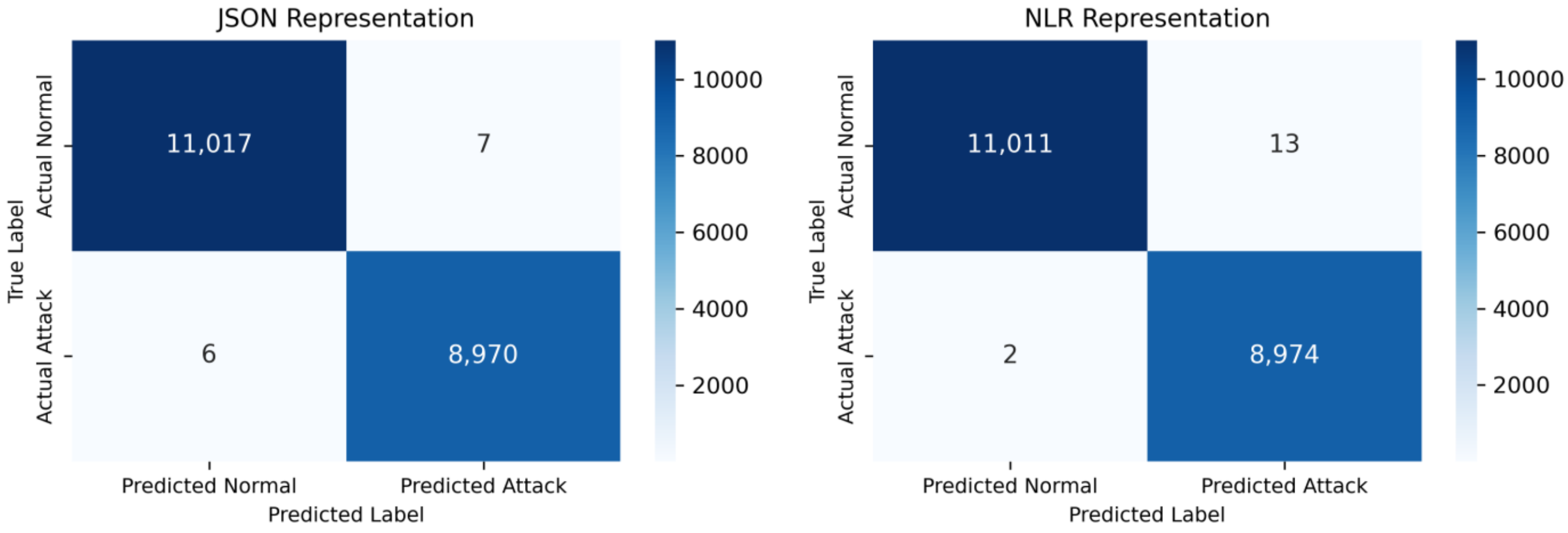

Figure 6: Confusion matrices of the proposed RAG-based framework (gemma-4-31b-IT) under different traffic representation strategies.

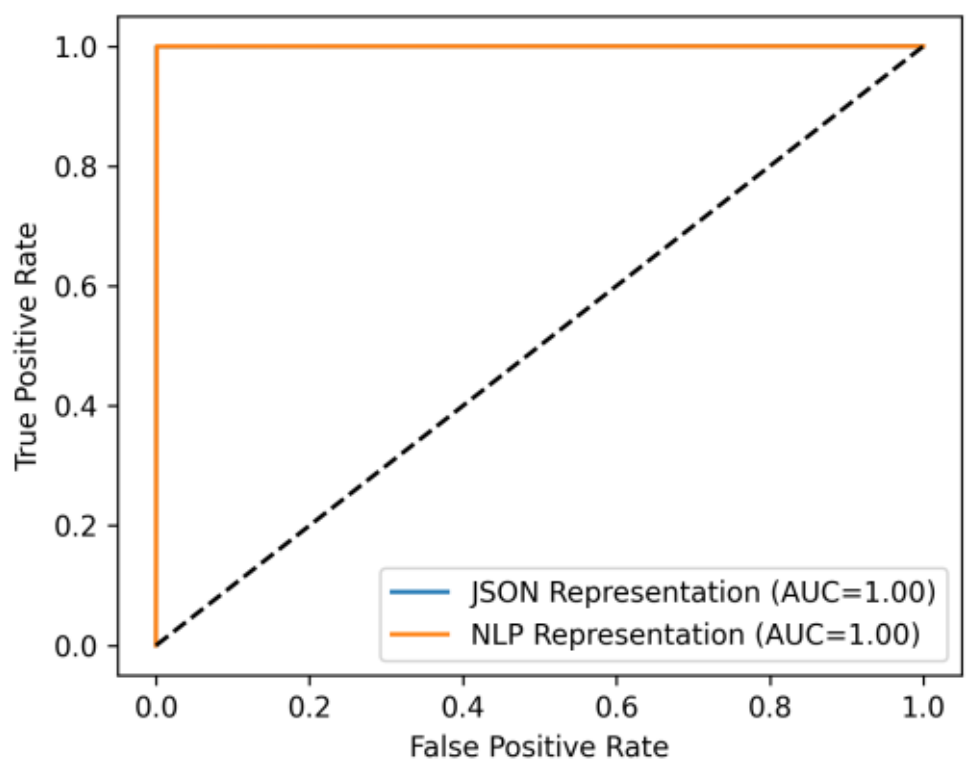

Figure 7: ROC curve results of the proposed RAG-based framework (gemma-4-31b-IT) under different traffic representation strategies.

### 4.5 Real-Time Detection and Mitigation Results

The real-time performance of the proposed RAG-based framework was evaluated under four Carpet-Bombing DDoS attack scenarios with different attack intensities to analyze its effectiveness in real-time attack detection and mitigation within SDN environments. Based on the comparative evaluation results presented in the previous sections, the structured JSON-based traffic representation was utilized during the real-time experiments due to its slightly superior and more stable detection performance compared with the natural language-based representation. During the evaluation process, several network behavior indicators were continuously monitored, including controller CPU utilization, Packet-In message rates, overall network traffic behavior, and traffic directed toward the victim servers. In addition, the real-time analysis compared the network behavior under three different operational conditions, namely attack-free operation, attack scenarios without mitigation, and attack scenarios with the proposed mitigation mechanism enabled. The obtained results demonstrate the effectiveness of the proposed framework in maintaining network stability and reducing the operational impact of Carpet-Bombing DDoS attacks under different attack intensities.

#### 4.5.1 Real-Time Detection and Mitigation Analysis of the First Scenario

The first real-time evaluation scenario analyzes the performance of the proposed RAG-based framework under a high-rate Carpet-Bombing DDoS attack, generating approximately 100,000 packets per second. As illustrated in Figure 8, the attack scenario without mitigation caused a substantial increase in SDN controller CPU utilization compared with the normal network condition, indicating the significant computational overhead generated by the distributed attack traffic. However, shortly after enabling the proposed mitigation mechanism, the CPU utilization noticeably decreased and gradually stabilized within a short period compared with the attack scenario without mitigation.

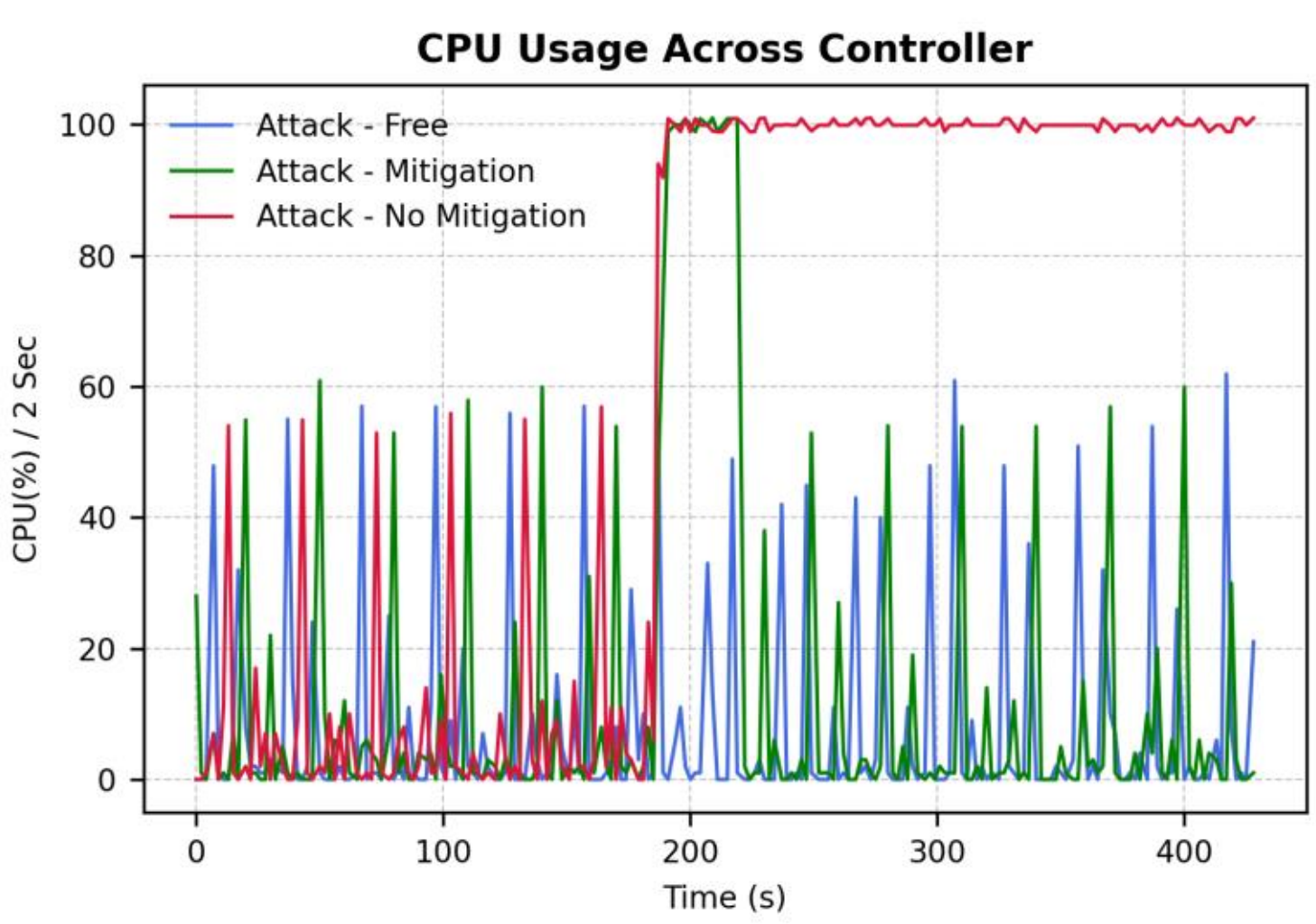

Figure 8: SDN controller CPU utilization under Carpet-Bombing DDoS attack in the first scenario.

Similarly, Figure 9 demonstrates a sharp increase in Packet-In message rates during the attack period without mitigation due to the large number of malicious traffic flows processed by the SDN controller. After activating the proposed mitigation mechanism, the Packet-In traffic was rapidly reduced within seconds of attack detection, indicating that the proposed framework successfully limited abnormal controller communication overhead.

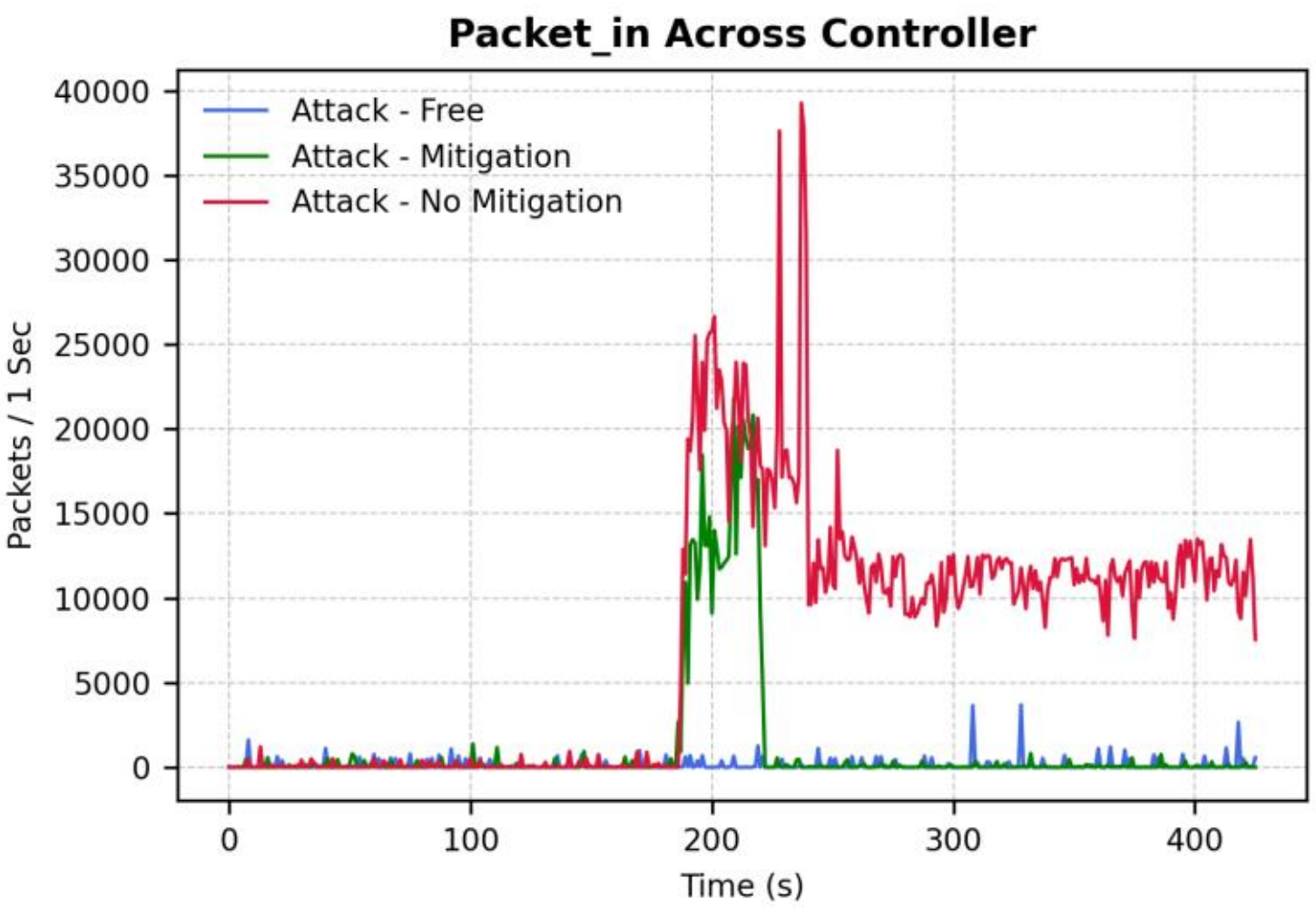

Figure 9: Packet-In message behavior under Carpet-Bombing DDoS attack in the first scenario.

Furthermore, the overall network traffic behavior presented in Figure 10 reveals severe network congestion and abnormal traffic growth during the attack scenario without mitigation. In contrast, the proposed mitigation mechanism effectively reduced excessive traffic pressure and improved overall network stability shortly after attack detection.

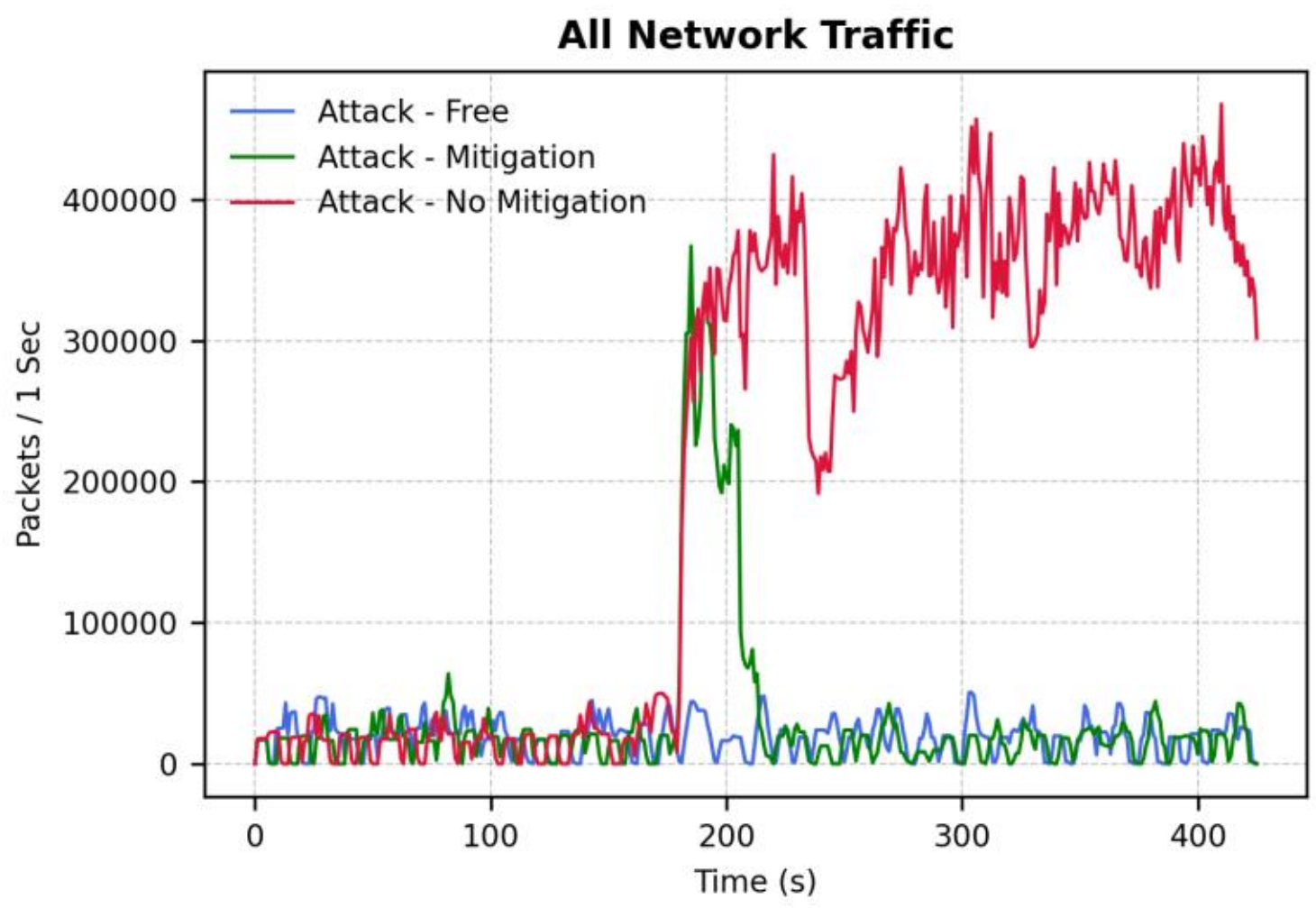

Figure 10: Overall network traffic behavior under Carpet-Bombing DDoS attack in the first scenario.

Finally, Figure 11 illustrates the traffic behavior directed toward the victim servers during the Carpet-Bombing DDoS attack. The results show that the malicious traffic was distributed across multiple targets simultaneously, making the traffic toward each individual victim appear less suspicious. However, when considering the attack on all servers collectively, the traffic becomes extremely dangerous, as illustrated in Figure 10. Nevertheless, the proposed framework successfully detected and mitigated the distributed attack behavior through contextual traffic analysis across the SDN environment within a short response time.

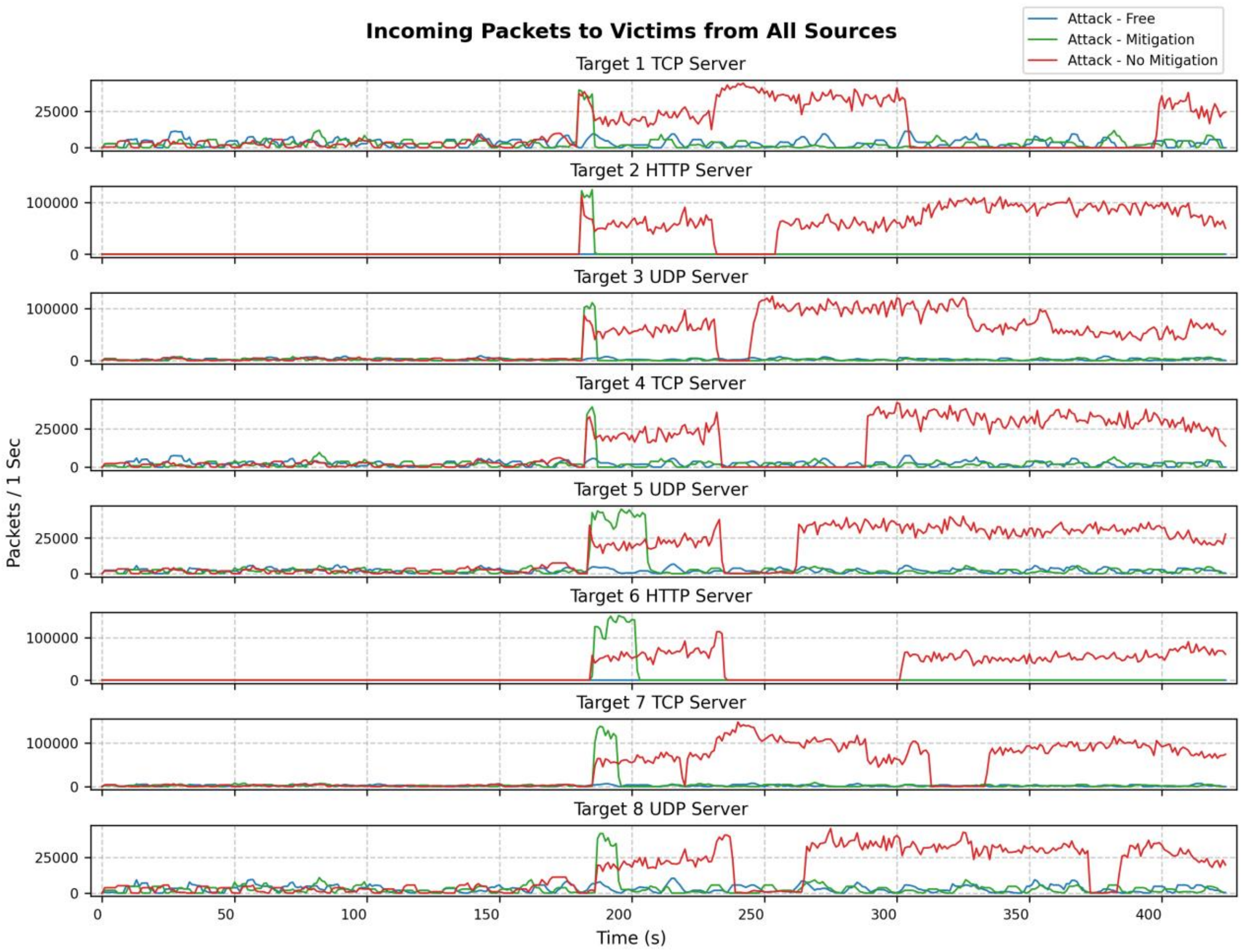

Figure 11: Traffic behavior directed toward victim servers under Carpet-Bombing DDoS attack in the first scenario.

### 4.5.2 Real-Time Detection and Mitigation Analysis of the Second Scenario

The second real-time evaluation scenario analyzes the performance of the proposed RAG-based framework under a Carpet-Bombing DDoS attack generating approximately 66,666 packets per second. As illustrated in Figure 12, the attack traffic still imposed considerable operational pressure on the SDN controller, resulting in a noticeable increase in CPU utilization during the attack period without mitigation. However, compared with the first scenario, the CPU fluctuation behavior appeared relatively lower due to the reduced attack intensity. After enabling the proposed mitigation mechanism, the controller utilization rapidly stabilized within a short response period.

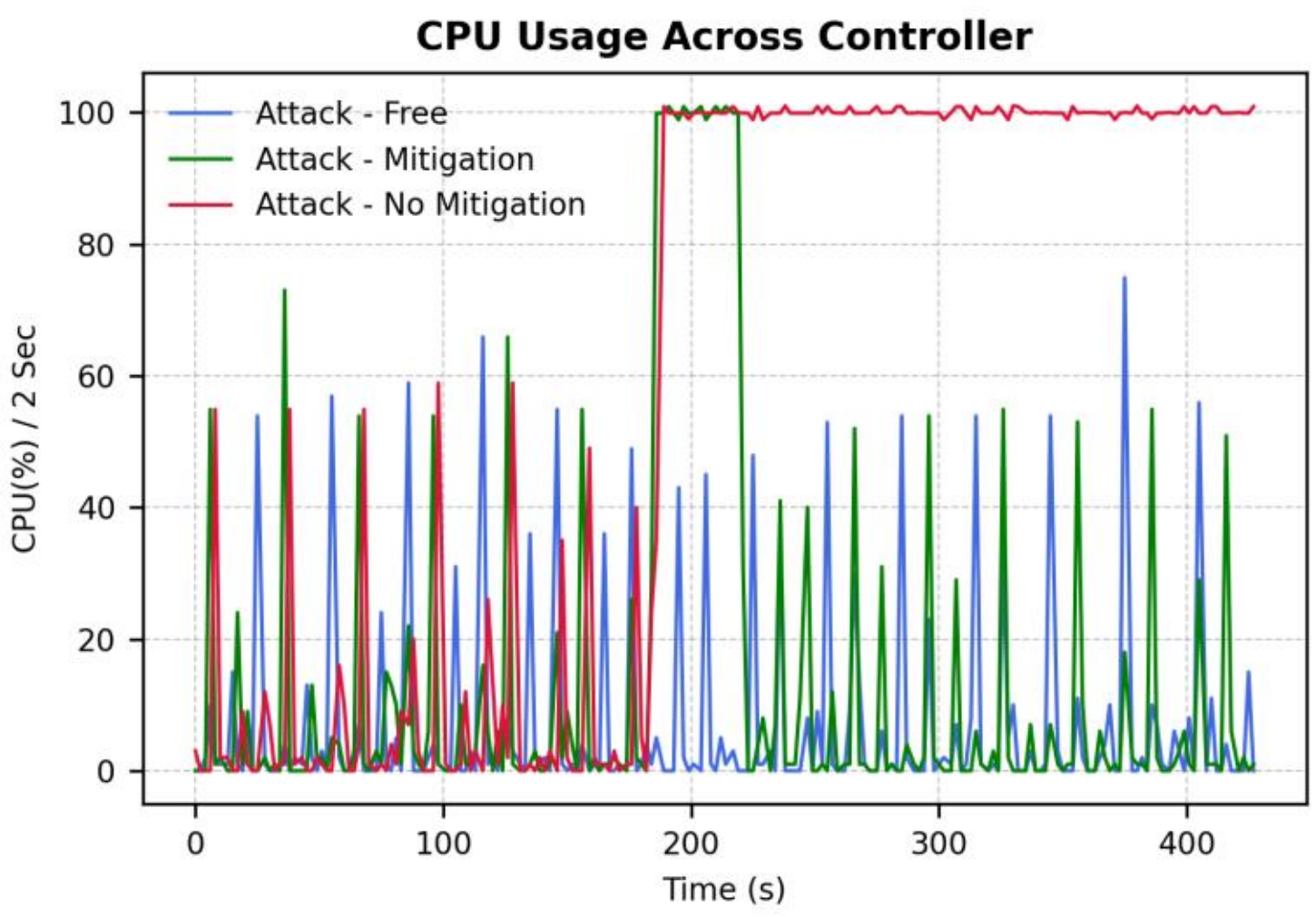

Figure 12: SDN controller CPU utilization under Carpet-Bombing DDoS attack in the second scenario.

Similarly, Figure 13 shows that the Packet-In message rate increased significantly during the attack period as a result of the abnormal malicious traffic flows processed by the controller. Nevertheless, the proposed mitigation mechanism successfully reduced the Packet-In traffic shortly after attack detection, preventing prolonged controller communication overload.

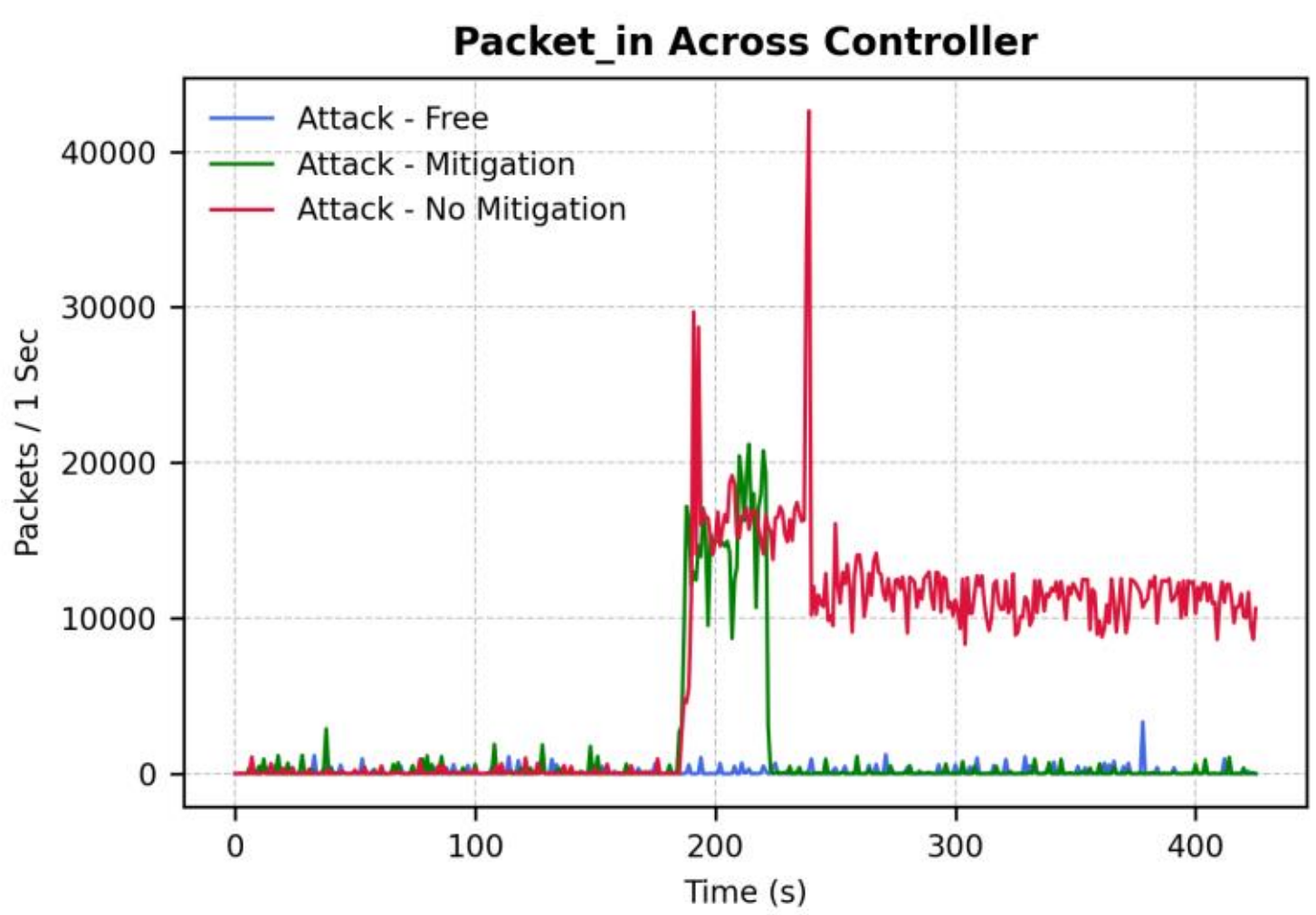

Figure 13: Packet-In message behavior under Carpet-Bombing DDoS attack in the second scenario.

Furthermore, the overall network traffic behavior presented in Figure 14 demonstrates that the attack continued to generate abnormal traffic activity and noticeable network congestion despite the lower attack rate compared with the first scenario. After mitigation activation, the abnormal traffic behavior was substantially reduced, resulting in improved network stability and traffic regulation.

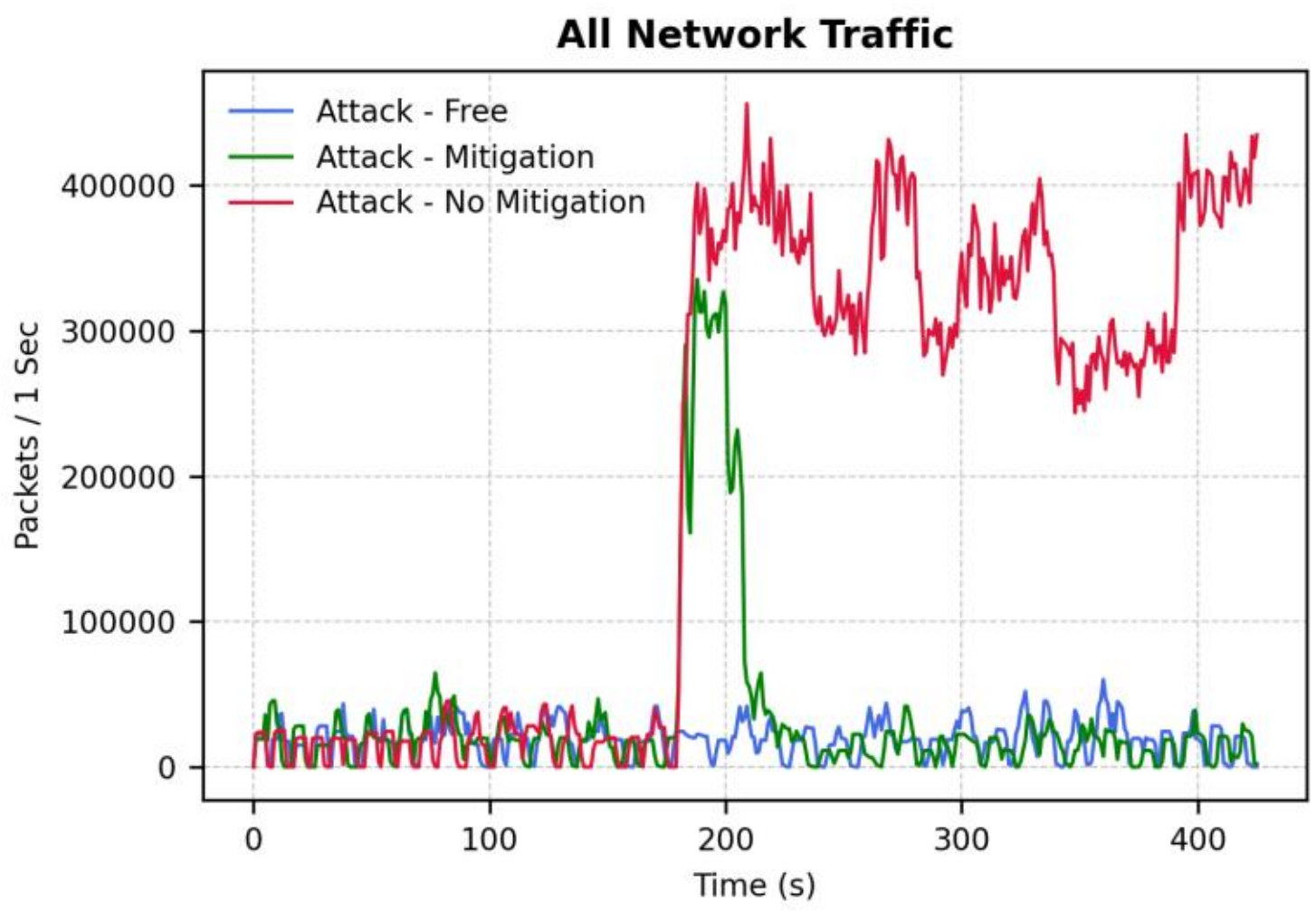

Figure 14: Overall network traffic behavior under Carpet-Bombing DDoS attack in the second scenario.

Finally, Figure 15 illustrates that the distributed attack traffic remained difficult to identify when analyzing each victim server independently. The results show that the malicious traffic was distributed across multiple targets simultaneously, making the traffic toward each individual victim appear less suspicious. However, when considering the collective traffic behavior across all targeted servers, the distributed attack impact becomes significantly more harmful, as illustrated in Figure 14. Overall, the proposed framework successfully maintained effective real-time detection and mitigation performance under moderate-high Carpet-Bombing DDoS attack conditions.

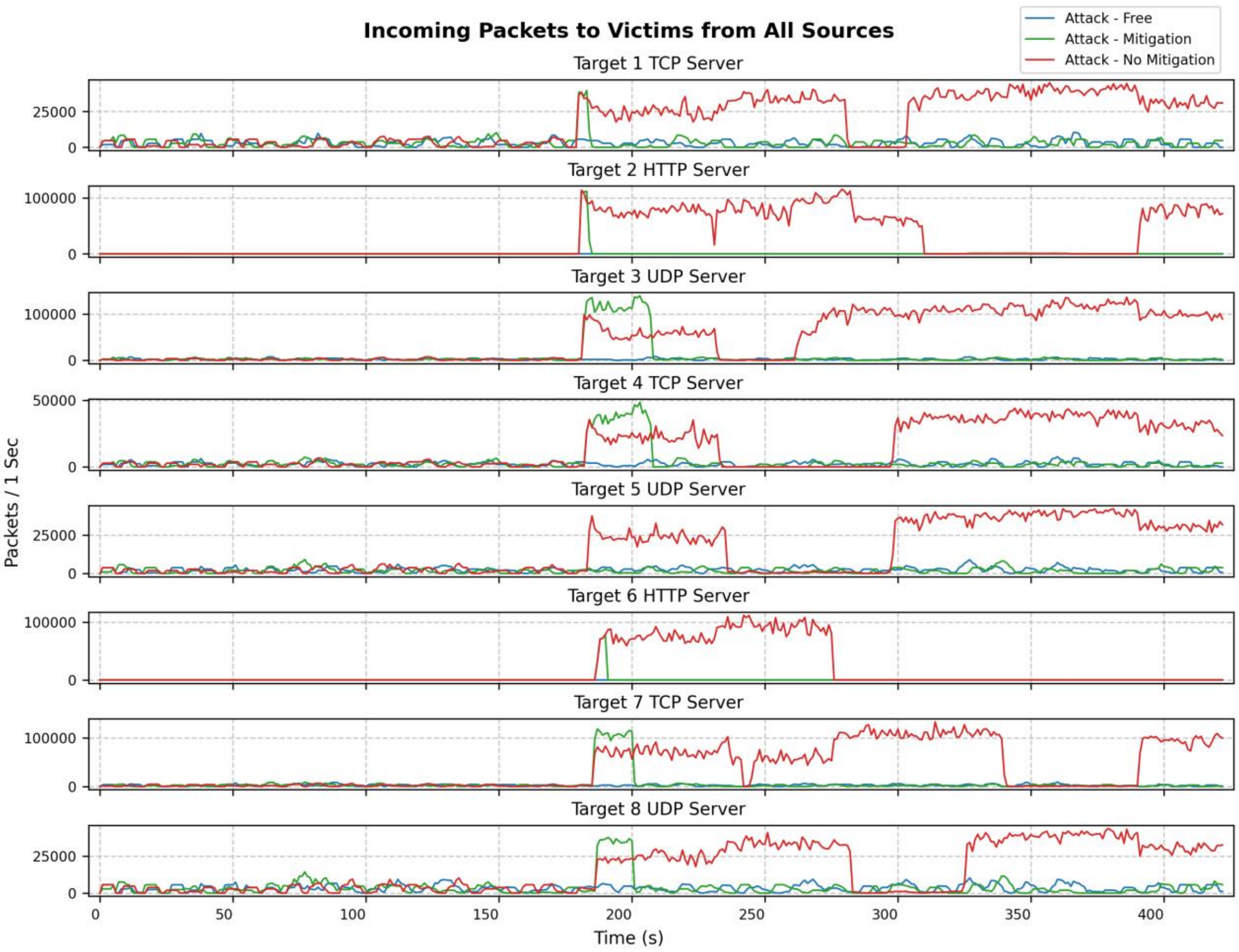

Figure 15: Traffic behavior directed toward victim servers under Carpet-Bombing DDoS attack in the second scenario.

### 4.5.3 Real-Time Detection and Mitigation Analysis of the Third Scenario

The third real-time evaluation scenario analyzes the performance of the proposed RAG-based framework under a Carpet-Bombing DDoS attack, generating approximately 40,000 packets per second. As illustrated in Figure 16, the attack scenario without mitigation continued to affect SDN controller stability by increasing CPU utilization compared with the normal network condition. However, the operational pressure on the controller became less severe compared with the previous attack scenarios due to the lower attack intensity. After enabling the proposed mitigation mechanism, the controller CPU utilization rapidly stabilized within a short response period.

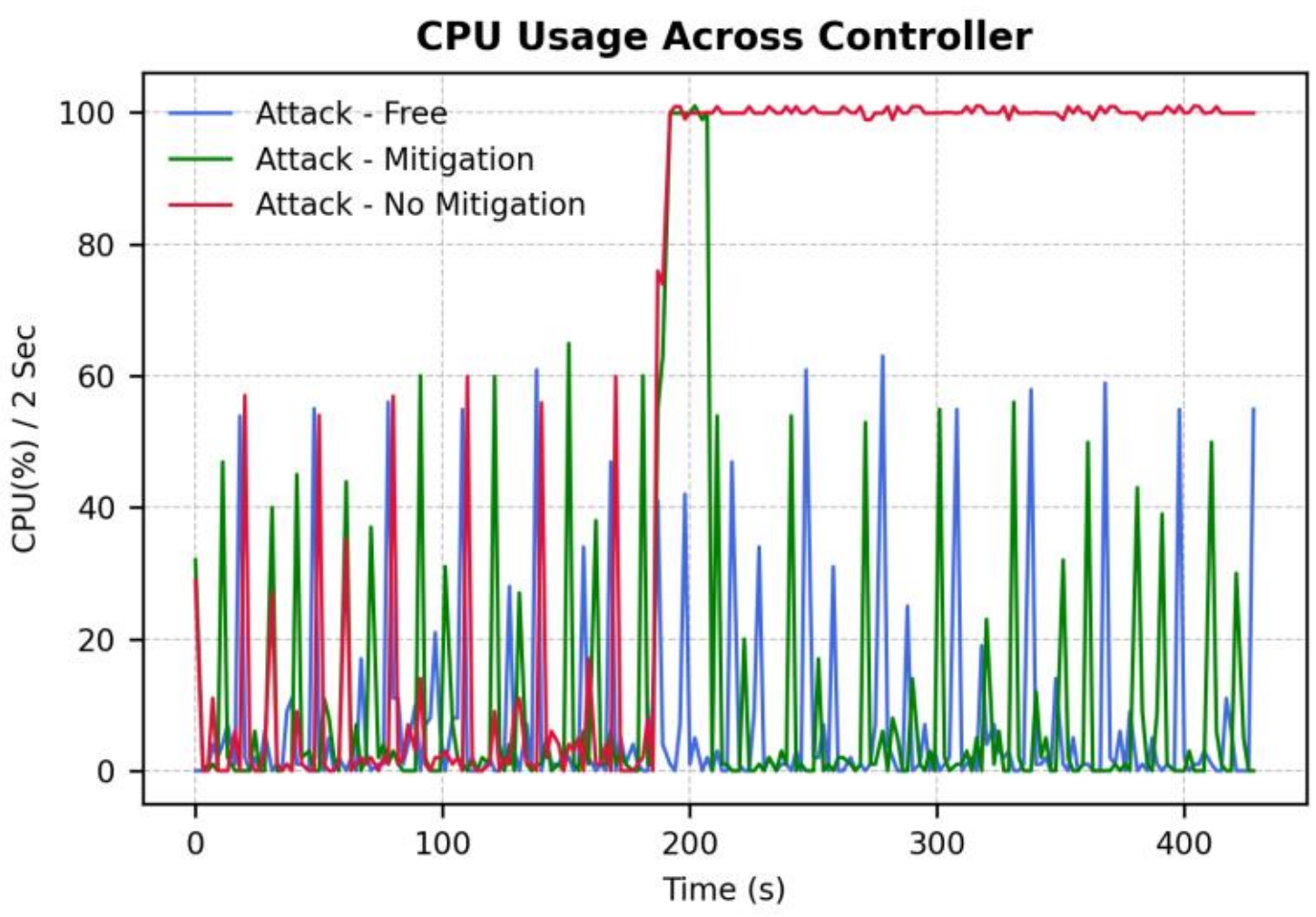

Figure 16: SDN controller CPU utilization under Carpet-Bombing DDoS attack in the third scenario.

Similarly, Figure 17 demonstrates that the Packet-In message rate still increased noticeably during the attack period without mitigation, reflecting the abnormal communication overhead introduced by the distributed malicious traffic flows. Nevertheless, the proposed mitigation mechanism successfully reduced the Packet-In traffic shortly after attack detection, maintaining more stable controller-switch communication behavior.

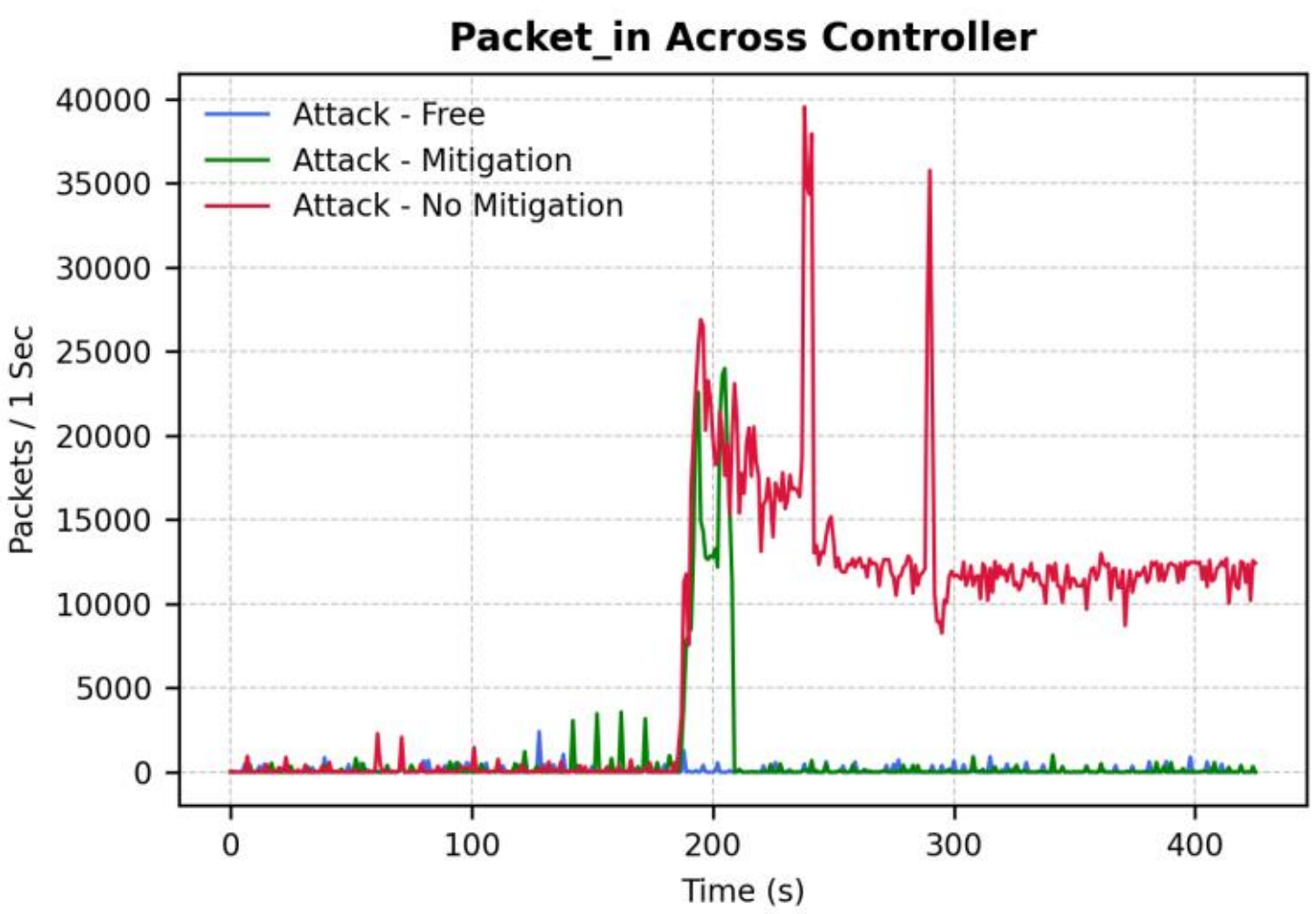

Figure 17: Packet-In message behavior under Carpet-Bombing DDoS attack in the third scenario.

Furthermore, the overall network traffic behavior illustrated in Figure 18 reveals that the distributed attack continued to generate abnormal traffic activity and network instability despite the moderate attack rate. After mitigation activation, the abnormal traffic behavior was significantly reduced, resulting in improved network stability and reduced traffic congestion across the SDN environment.

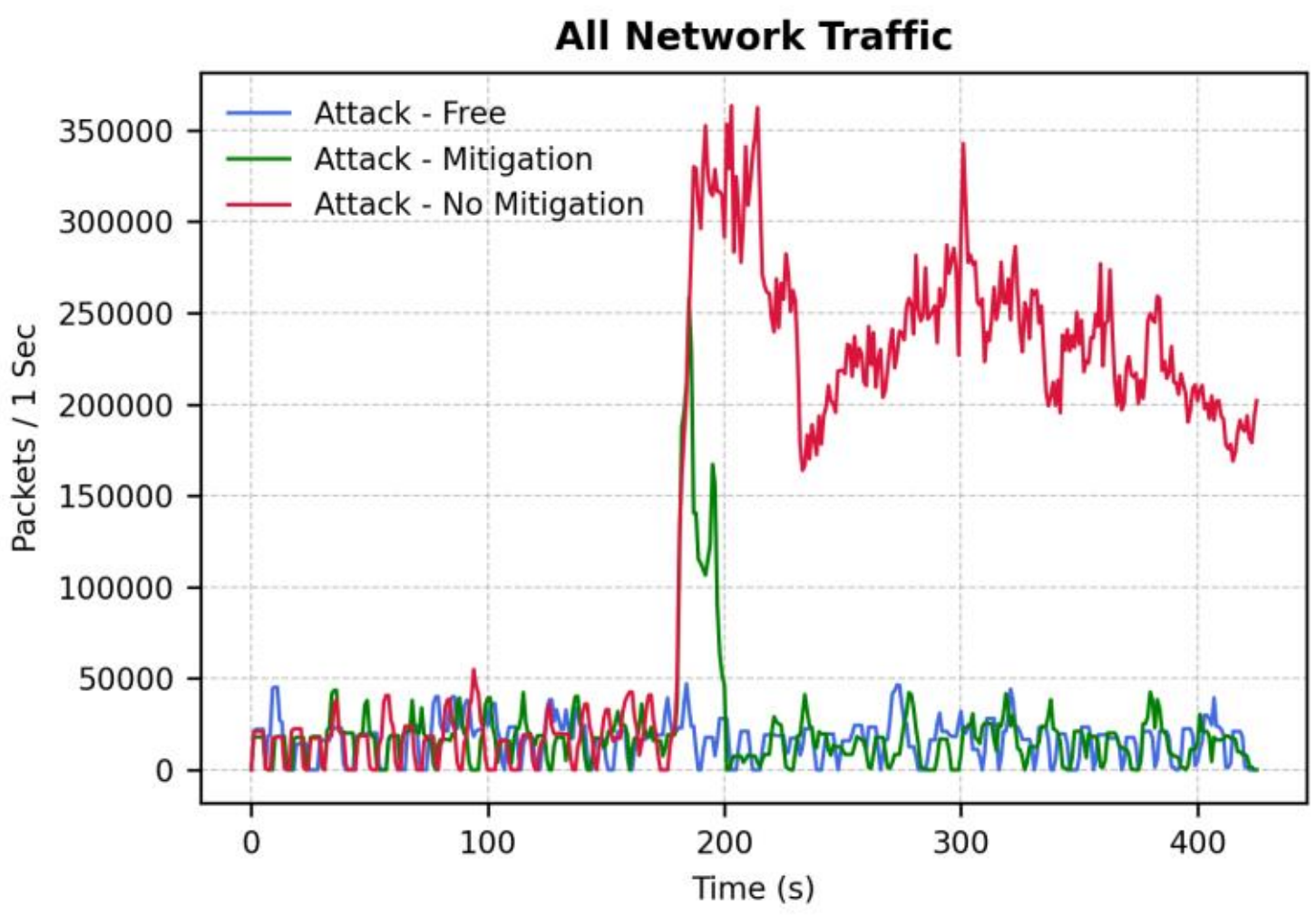

Figure 18: Overall network traffic behavior under Carpet-Bombing DDoS attack in the third scenario.

Finally, Figure 19 illustrates the traffic behavior directed toward the victim servers during the Carpet-Bombing DDoS attack. The results show that the malicious traffic was distributed across multiple targets simultaneously, making the traffic toward each individual victim appear less suspicious. However, when considering the attack traffic collectively across all targeted servers, the overall network impact becomes significantly more dangerous, as illustrated in Figure 18. Overall, the proposed framework successfully maintained effective real-time detection and mitigation performance under moderate Carpet-Bombing attack conditions.

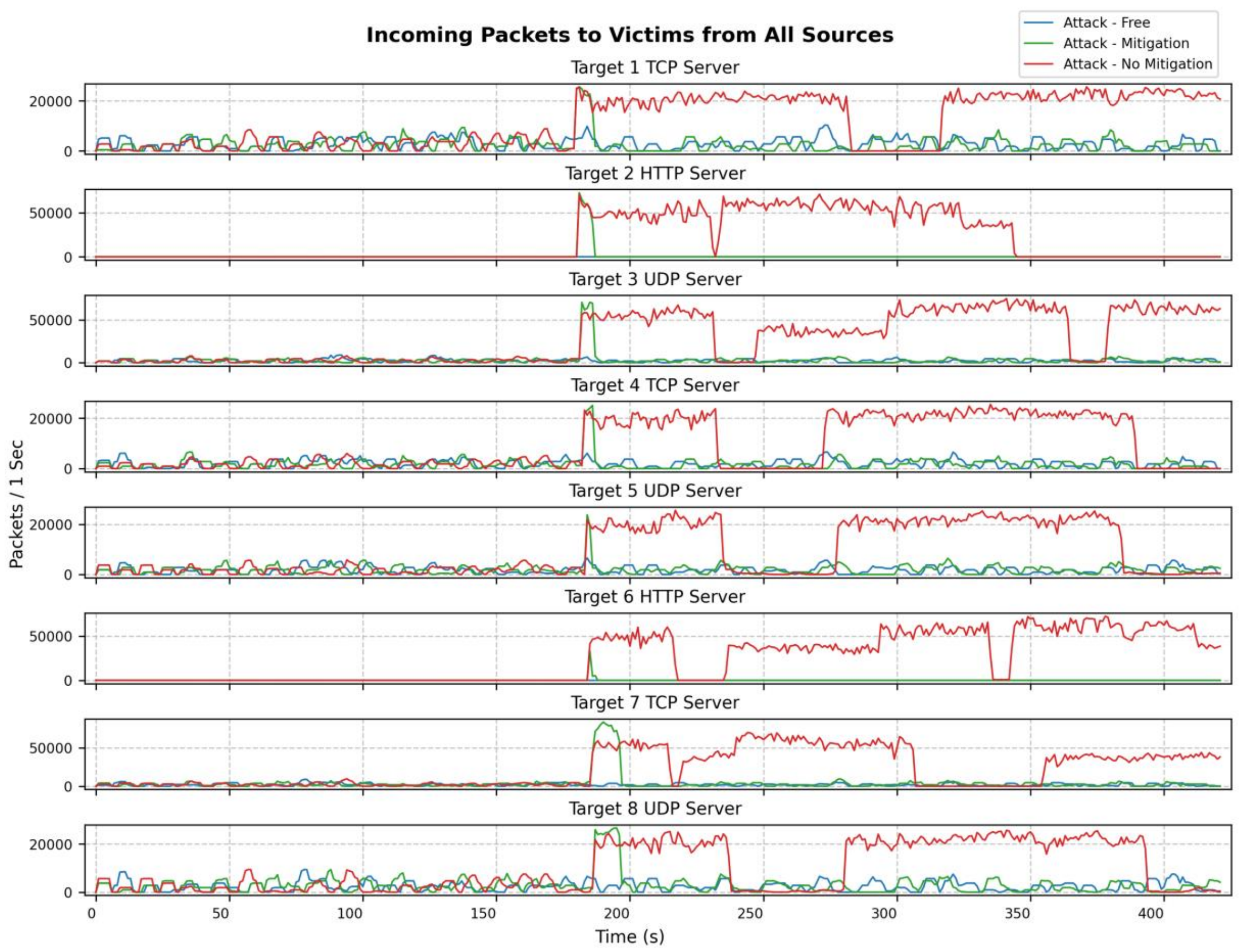

Figure 19: Traffic behavior directed toward victim servers under Carpet-Bombing DDoS attack in the third scenario.

### 4.5.4 Real-Time Detection and Mitigation Analysis of the Fourth Scenario

The fourth real-time evaluation scenario analyzes the performance of the proposed RAG-based framework under a low-rate Carpet-Bombing DDoS attack generating approximately 20,000 packets per second. As illustrated in Figure 20, the attack scenario without mitigation still produced noticeable effects on SDN controller operation, resulting in increases in CPU utilization compared with the normal network condition. Although the operational impact was lower than the previous scenarios, the obtained results demonstrate that even lower-rate distributed attacks can still influence SDN controller stability if mitigation mechanisms are not applied. However, after enabling the proposed mitigation mechanism, the controller utilization rapidly stabilized within a short response period.

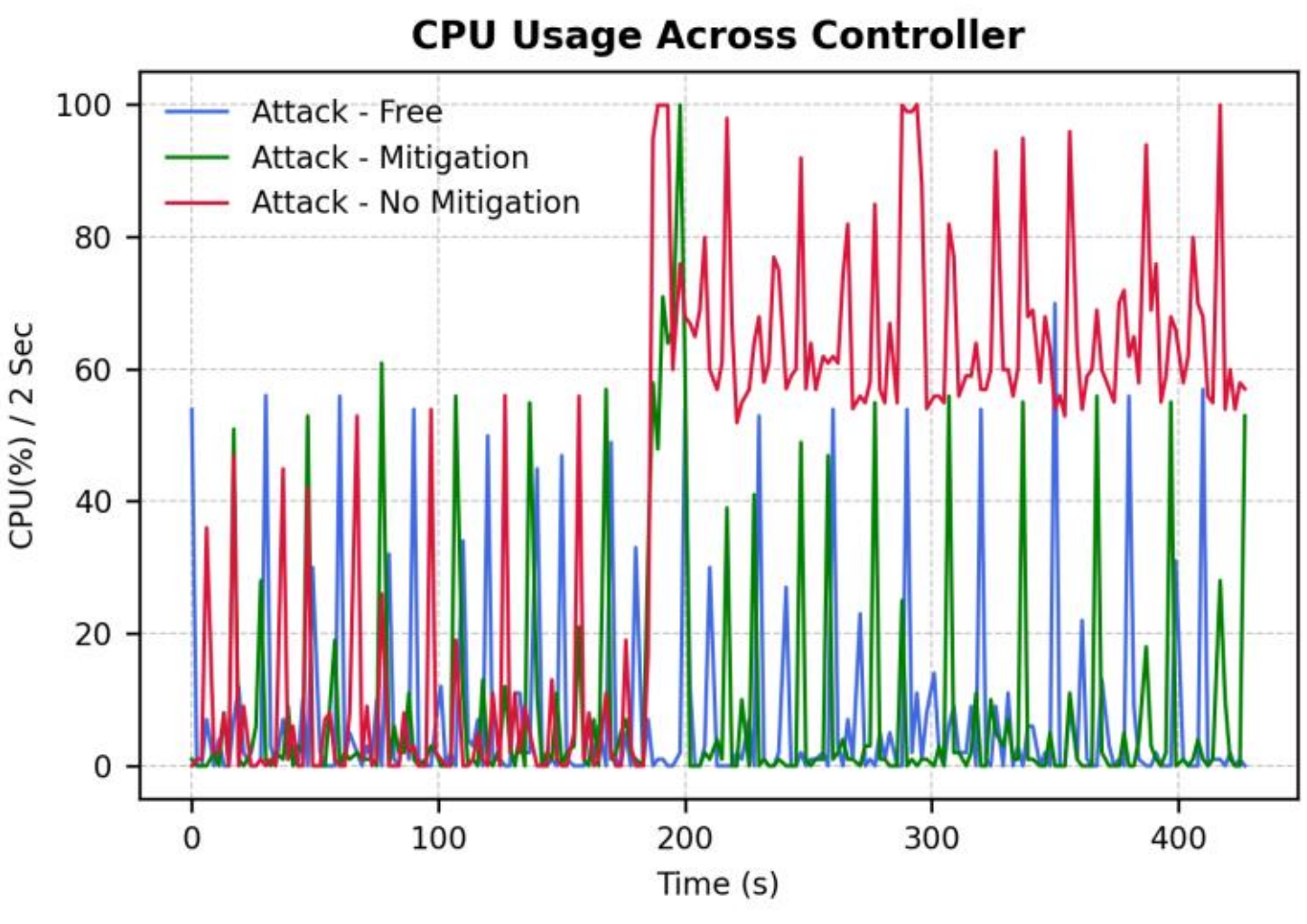

Figure 20: SDN controller CPU utilization under Carpet-Bombing DDoS attack in the fourth scenario.

Similarly, Figure 21 shows that the Packet-In message rate increased during the attack period due to the abnormal malicious traffic flows processed by the SDN controller. After activating the proposed mitigation mechanism, the Packet-In traffic was effectively reduced shortly after attack detection, maintaining more stable controller communication behavior.

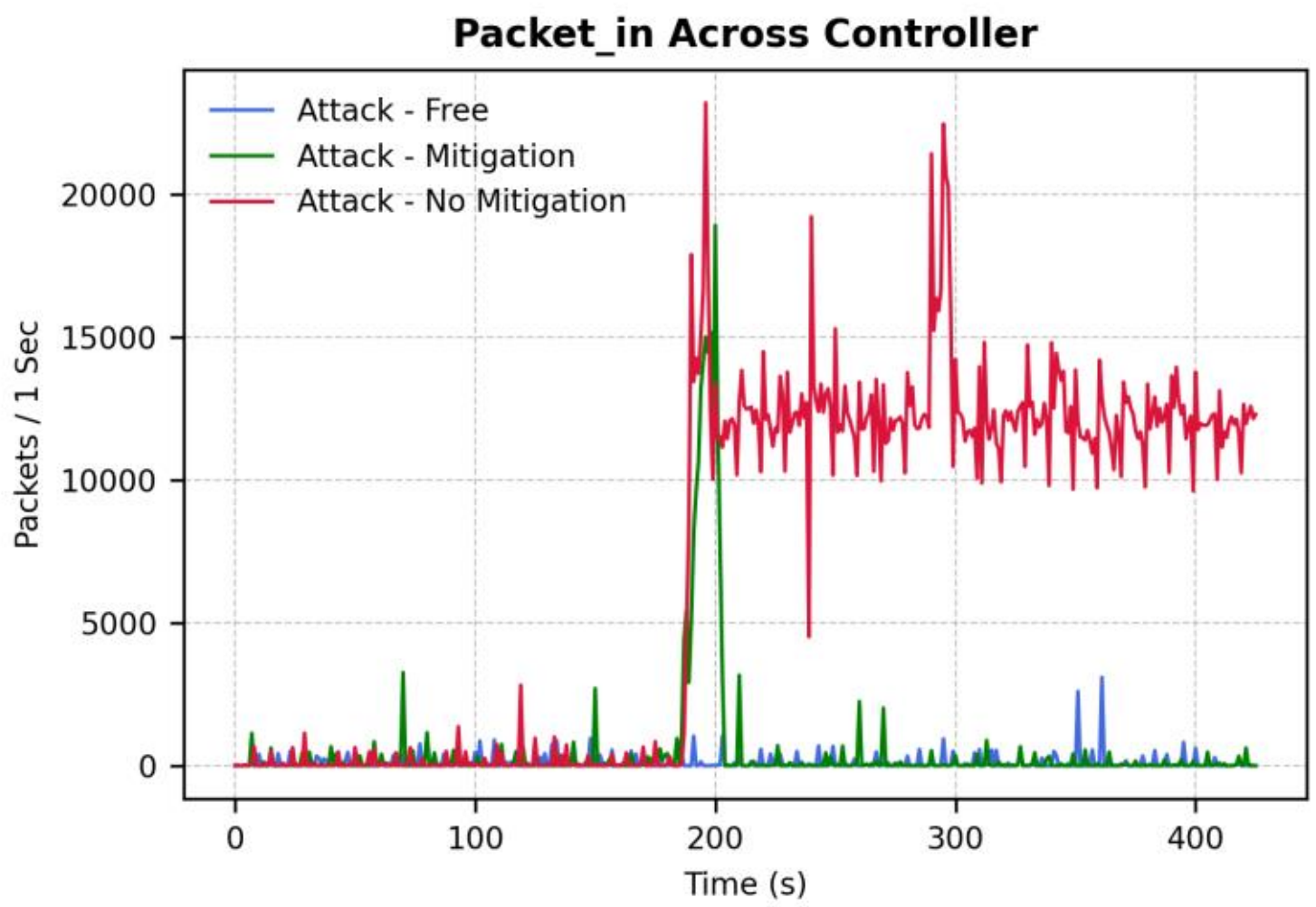

Figure 21: Packet-In message behavior under Carpet-Bombing DDoS attack in the fourth scenario.

Furthermore, the overall network traffic behavior illustrated in Figure 22 demonstrates that the distributed attack continued to generate abnormal traffic activity despite the lower attack intensity. Nevertheless, the proposed mitigation mechanism successfully reduced excessive traffic behavior and improved overall network stability after mitigation activation.

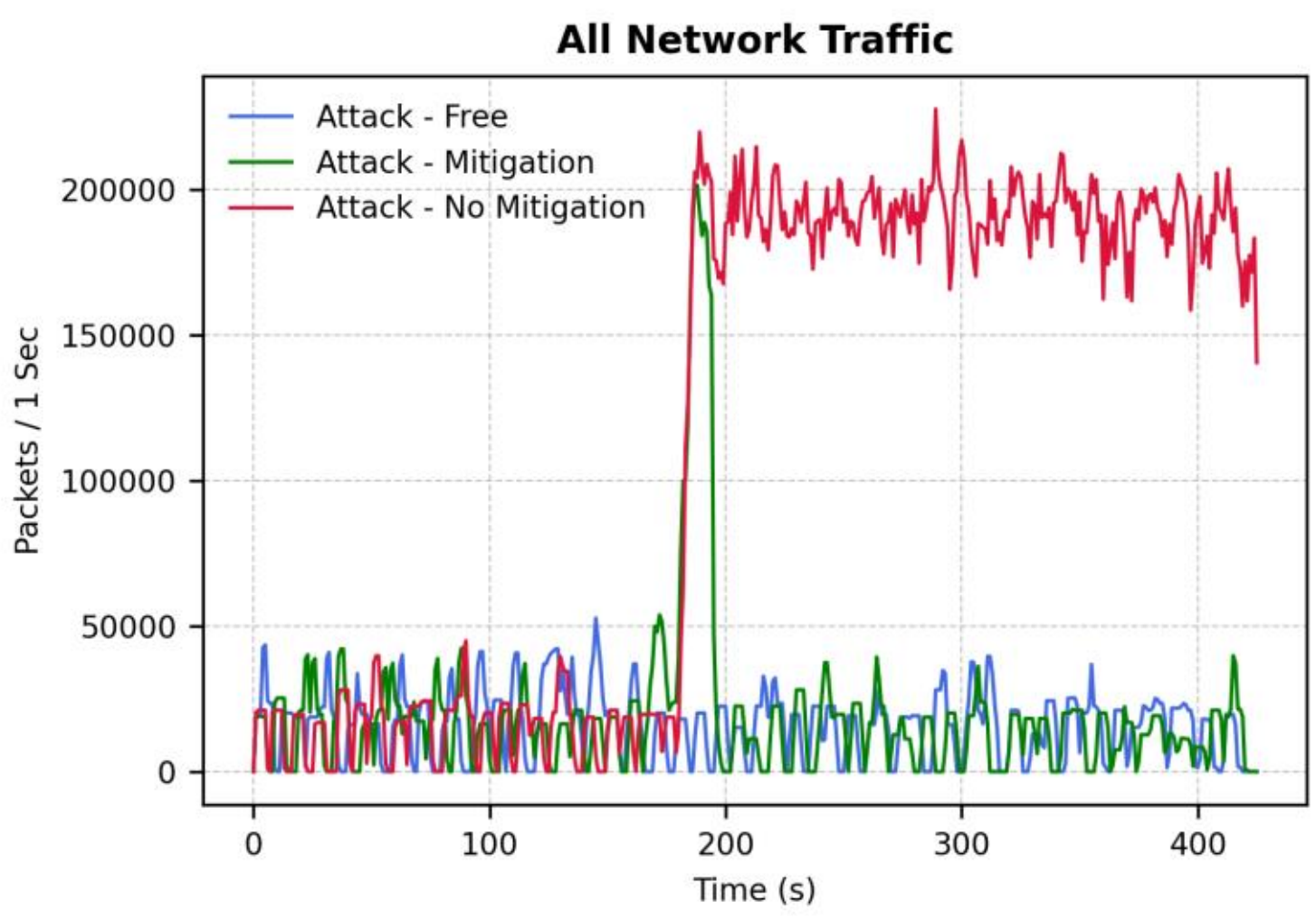

Figure 22: Overall network traffic behavior under Carpet-Bombing DDoS attack in the fourth scenario.

Finally, Figure 23 illustrates the traffic behavior directed toward the victim servers during the Carpet-Bombing DDoS attack. The results show that the malicious traffic was distributed across multiple targets simultaneously, making the traffic toward each individual victim appear less suspicious. However, when analyzing the attack collectively across all targeted servers, the distributed attack impact becomes significantly more dangerous, as illustrated in Figure 22. Overall, the proposed framework successfully maintained effective real-time detection and mitigation performance even under lower-rate Carpet-Bombing DDoS attack conditions.

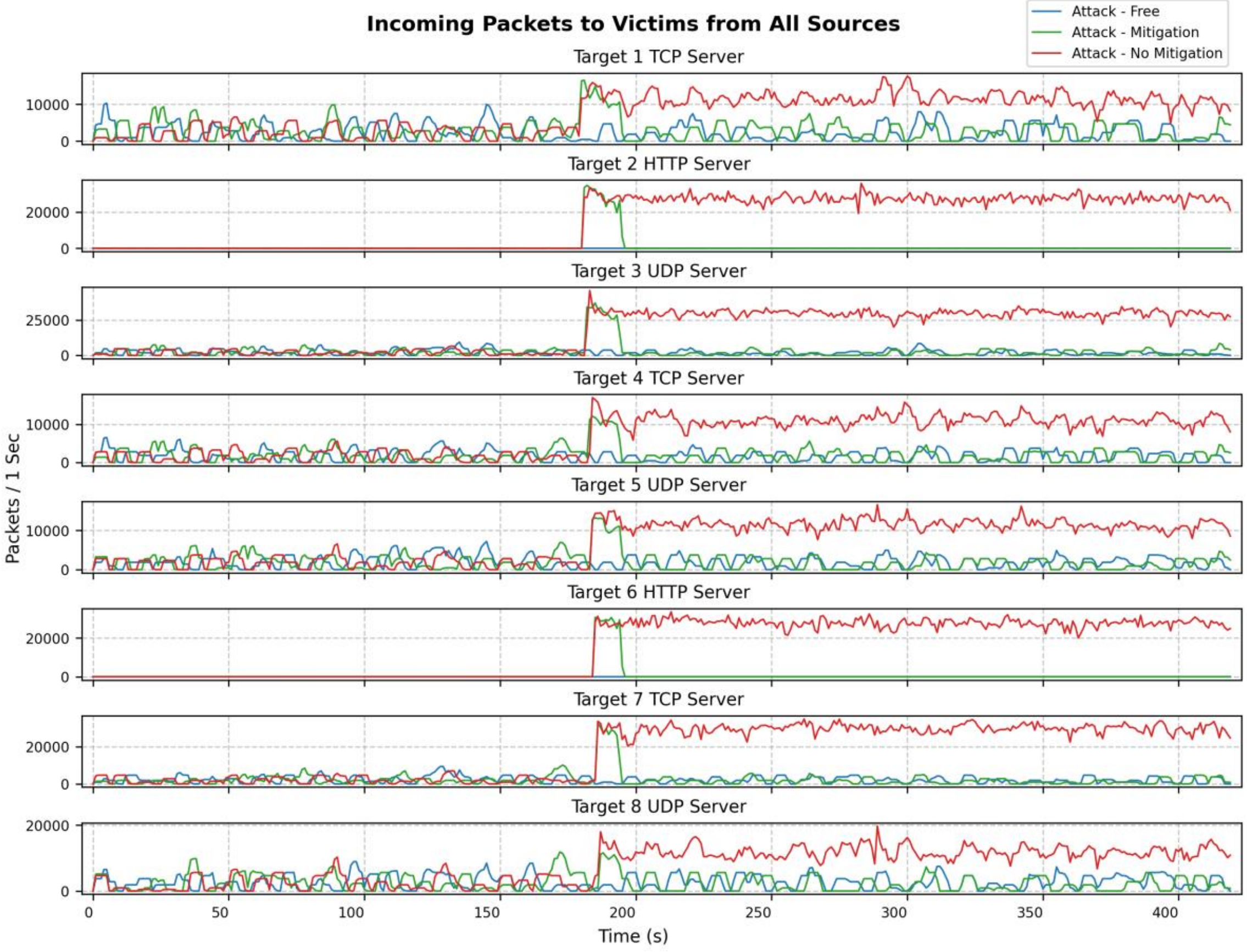

Figure 23: Traffic behavior directed toward victim servers under Carpet-Bombing DDoS attack in the fourth scenario.

### 4.6 Comparative Evaluation Using Different Large Language Models

To further analyze the effectiveness of the proposed RAG-based framework, a comparative evaluation was conducted using multiple state-of-the-art LLMs under the two proposed traffic representation strategies, namely the structured JSON-based representation and the natural language-based representation. The proposed framework primarily utilizes the Gemma-4-31B-IT model as the core inference component for contextual traffic analysis and attack classification. Additional LLMs, including Grok-4.1-fast, DeepSeek-v3.2, Gemini-2.5-Flash, Llama-3.1-70B-Instruct, and GPT-4o-mini, were investigated to compare their detection performance and inference behavior within the proposed retrieval-based framework. Several performance evaluation metrics, including Accuracy, Precision, Recall, F1-Score, average request processing time, confusion matrices, and ROC-AUC analysis, were utilized to comprehensively evaluate the detection capability of the investigated LLMs under both traffic representation strategies.

#### 4.6.1 Evaluation Using Structured JSON Representation

The structured JSON-based traffic representation was evaluated using multiple state-of-the-art LLMs within the proposed RAG-based framework to compare their detection effectiveness with the proposed framework configuration based on the Gemma-4-31B-IT model for detecting Carpet-Bombing DDoS attacks in SDN environments. The obtained results demonstrate that the proposed framework achieved highly accurate and stable traffic classification performance under the structured JSON representation across all investigated LLMs. In addition, the experimental results indicate that the proposed framework configuration utilizing the Gemma-4-31B-IT model as the core inference component achieved the strongest overall detection performance compared with the other investigated LLMs. The structured JSON representation also provided organized contextual traffic information that facilitated more consistent retrieval behavior and effective inference operations during attack classification. Table 4 summarizes the overall performance results, while Figures 24 and 25 illustrate the corresponding confusion matrices and ROC curve results obtained during the evaluation process.

Table 4: Comparative performance results of different LLMs under the structured JSON-based traffic representation.

| Model | Accuracy | Precision | Recall | F1-score | Average Request Time |
|---|---|---|---|---|---|
| Gemma-4-31B-IT | 99.94% | 99.92% | 99.93% | 99.93% | 2.10/s |
| Grok-4.1-fast | 99.88% | 99.83% | 99.90% | 99.87% | 3.49/s |
| DeepSeek-v3.2 | 99.83% | 99.64% | 99.97% | 99.81% | 2.47/s |
| Gemini-2.5-Flash | 99.71% | 99.72% | 99.63% | 99.68% | 1.04/s |
| Llama-3.1-70B-Instruct | 99.55% | 99.00% | 100% | 99.50% | 1.14/s |
| GPT-4o-mini | 99.34% | 98.71% | 99.82% | 99.26% | 0.95/s |

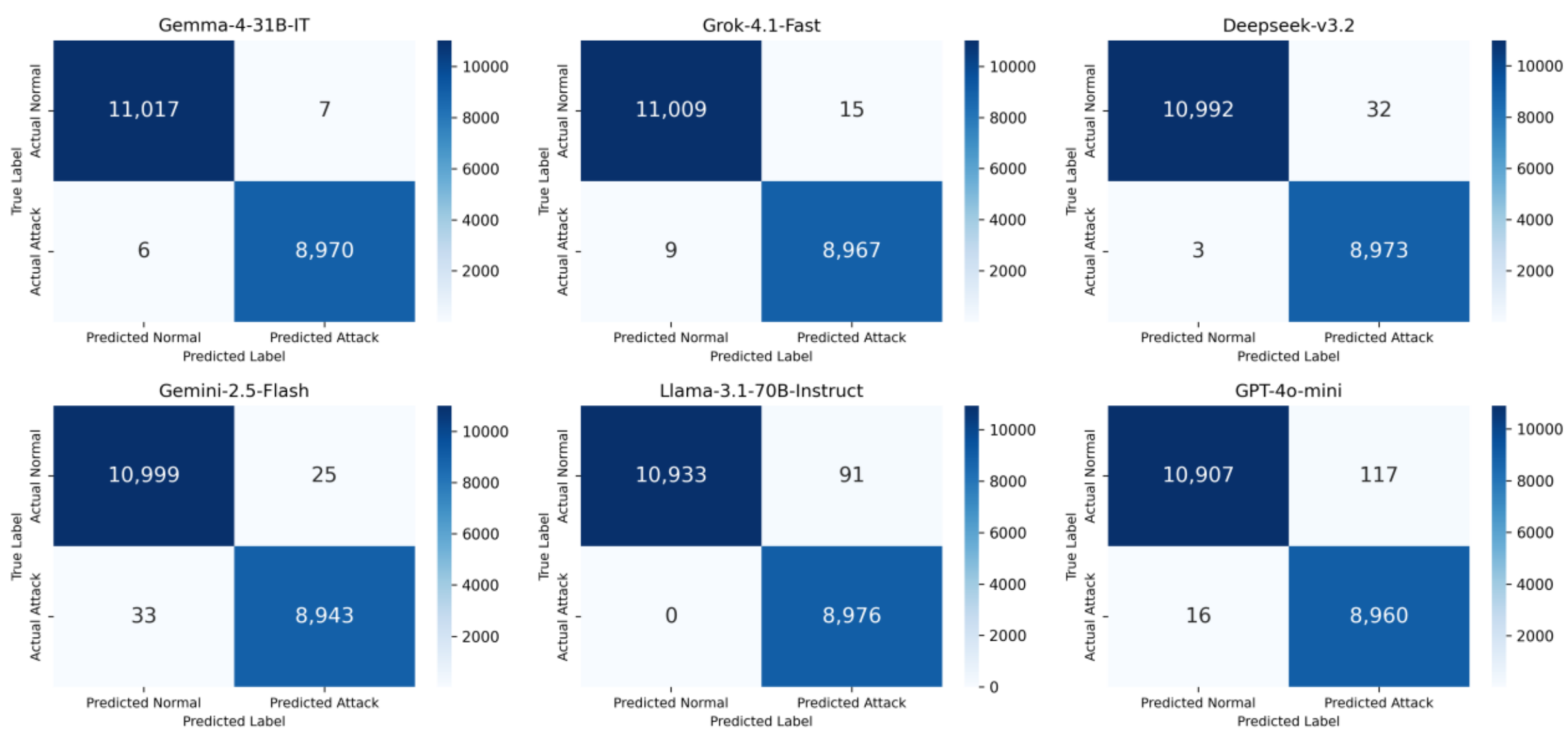

Figure 24: Confusion matrices of different LLMs under the structured JSON-based traffic representation.

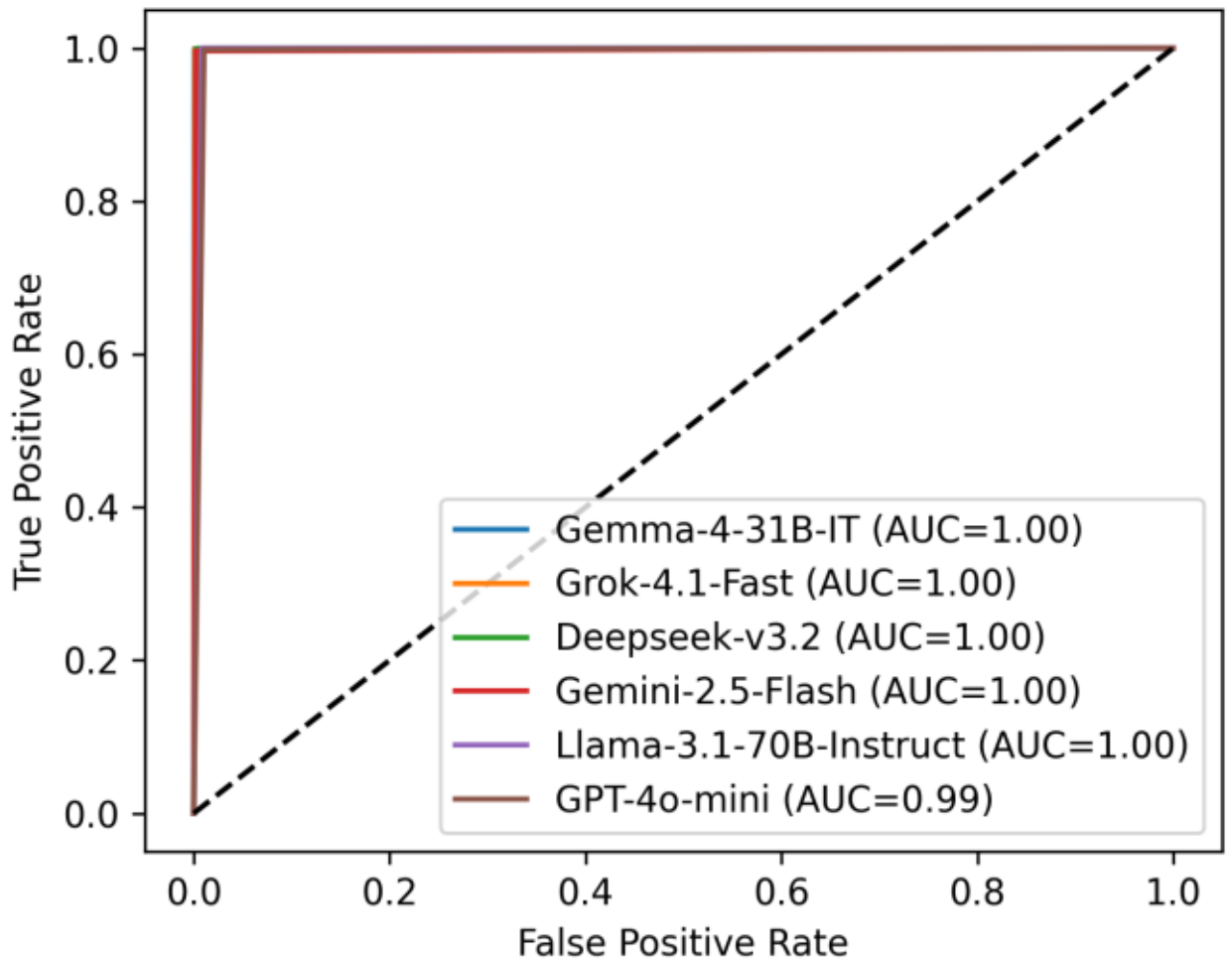

Figure 25: ROC curve results of different LLMs under the structured JSON-based traffic representation.

Table 4 demonstrates that the proposed RAG-based framework achieved excellent detection performance under the structured JSON-based traffic representation across all investigated LLMs. The proposed framework configuration based on Gemma-4-31B-IT achieved the strongest overall performance, obtaining an accuracy of 99.94%, precision of 99.92%, recall of 99.93%, and F1-score of 99.93%, while maintaining an average request processing time of approximately 2 seconds. Similarly, Grok-4.1-fast and DeepSeek-v3.2 also achieved highly reliable detection performance with accuracy values exceeding 99.80%, demonstrating the effectiveness of the proposed retrieval-based framework under different LLM architectures. In contrast, GPT-4o-mini achieved the lowest overall classification performance among the investigated models, particularly in terms of false positive and false negative predictions, although it still maintained strong overall detection capability with an accuracy exceeding 99%. Furthermore, the obtained confusion matrices indicate that the structured JSON representation enabled highly stable classification behavior with very limited misclassification rates across most investigated models. The ROC curve analysis additionally demonstrates strong discrimination capability under the structured JSON representation, where most investigated LLMs achieved near-perfect AUC values close to 1.00. Overall, the obtained results indicate that the structured JSON-based traffic representation provides highly organized contextual traffic information that enhances retrieval consistency and supports effective LLM-based traffic classification within the proposed RAG-based framework.

### 4.6.2 Evaluation Using Natural Language-based Representation

The natural language-based traffic representation was evaluated using multiple state-of-the-art LLMs within the proposed RAG-based framework to compare their detection effectiveness with the proposed framework configuration based on the Gemma-4-31B-IT model for detecting Carpet-Bombing DDoS attacks in SDN environments. The obtained results demonstrate that the proposed framework maintained highly accurate traffic classification performance under the natural language-based traffic representation across all investigated LLMs. In addition, the experimental results indicate that the proposed framework configuration utilizing the Gemma-4-31B-IT model as the core inference component continued to achieve the strongest overall detection performance compared with the other investigated LLMs. The natural language-based traffic representation also enabled effective contextual traffic analysis by transforming SDN interface statistics into descriptive textual traffic statements that could be processed during retrieval and inference operations. Table 5 summarizes the overall performance results, while Figures 26 and 27 illustrate the corresponding confusion matrices and ROC curve results obtained during the evaluation process.

Table 5: Comparative performance results of different LLMs under the natural language-based traffic representation.

| Model | Accuracy | Precision | Recall | F1-score | Average Request Time |
|---|---|---|---|---|---|
| Gemma-4-31B-IT | 99.93% | 99.86% | 99.98% | 99.92% | 2.08/s |
| Gemini-2.5-Flash | 99.81% | 99.58% | 100% | 99.78% | 1.22/s |
| Grok-4.1-fast | 99.79% | 99.61% | 99.92% | 99.77% | 4.77/s |
| DeepSeek-v3.2 | 99.63% | 99.27% | 99.91% | 99.59% | 2.41/s |
| Llama-3.1-70B-Instruct | 98.91% | 98.12% | 99.47% | 98.79% | 1.52/s |
| GPT-4o-mini | 98.56% | 97.00% | 99.87% | 98.41% | 0.92/s |

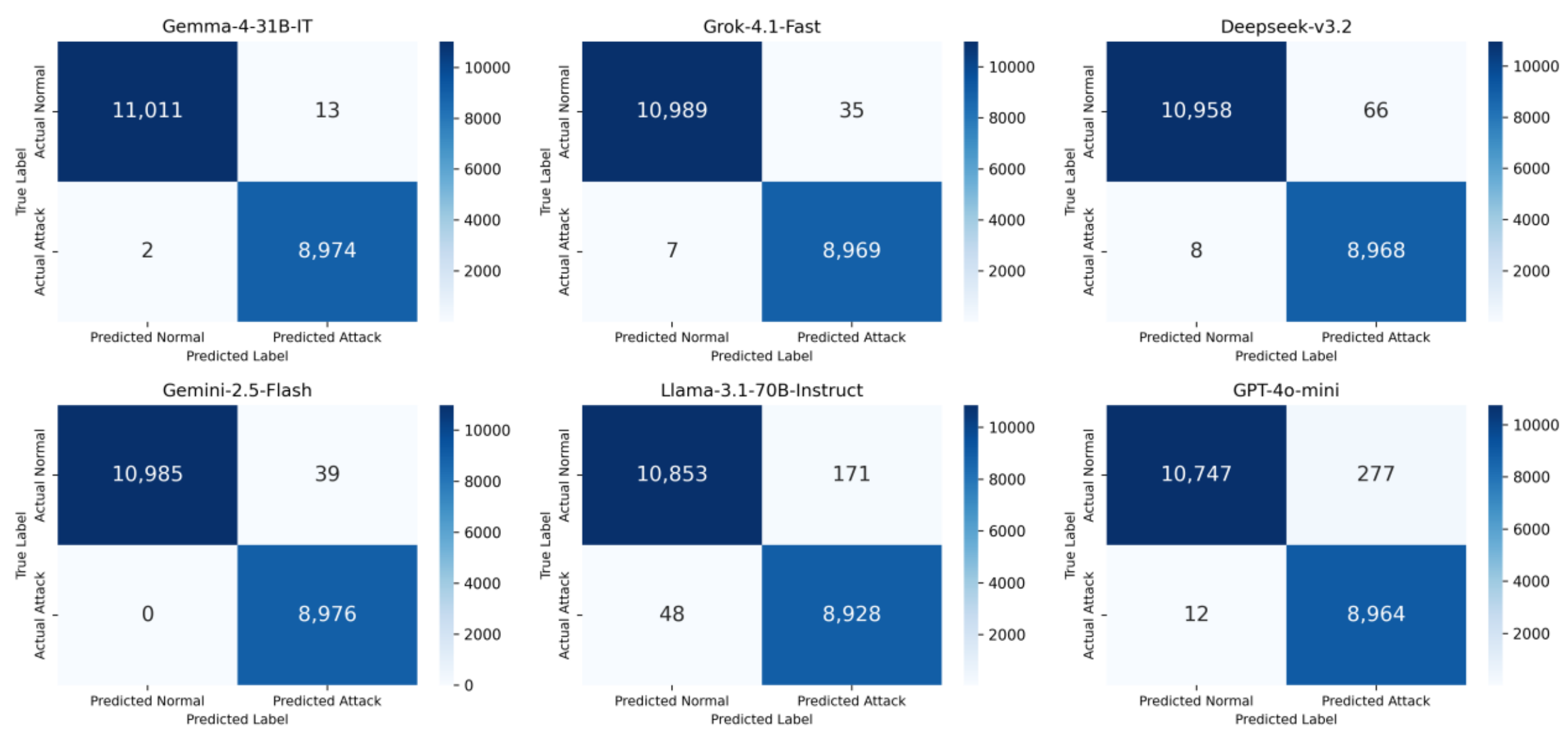

Figure 26: Confusion matrices of different LLMs using the natural language-based traffic representation.

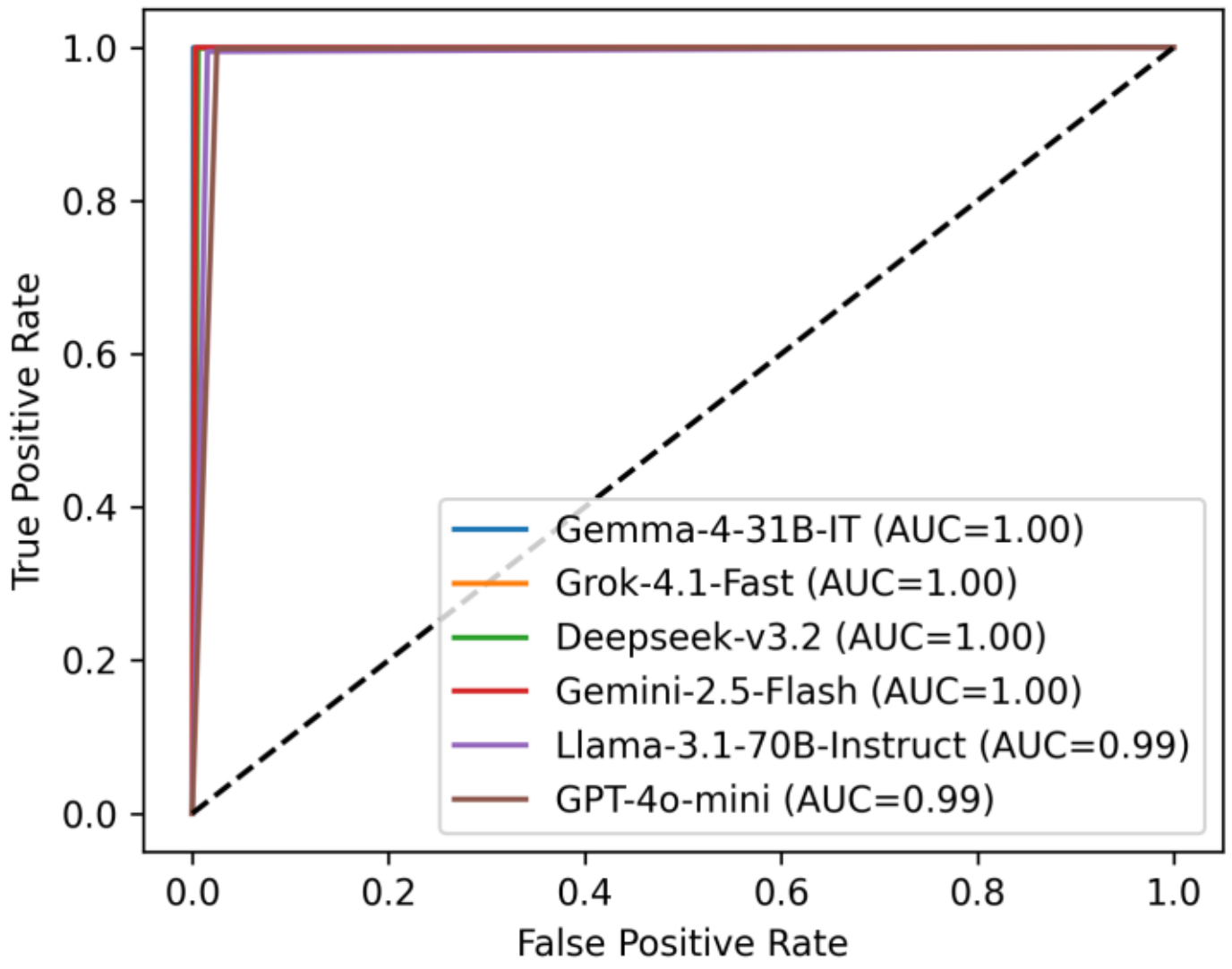

Figure 27: ROC-AUC analysis of different LLMs using the natural language-based traffic representation.

Table 5 demonstrates that the proposed RAG-based framework maintained highly accurate detection performance under the natural language-based traffic representation across all investigated LLMs. The proposed framework configuration based on Gemma-4-31B-IT achieved the strongest overall performance, obtaining an accuracy of 99.93%, precision of 99.86%, recall of 99.98%, and F1-score of 99.92%, while maintaining an average request processing time of approximately 2seconds. Similarly, Gemini-2.5-Flash, DeepSeek-v3.2, and Grok-4.1-fast also achieved highly reliable traffic classification performance with accuracy values exceeding 99%, demonstrating the effectiveness of the proposed retrieval-based framework under the natural language-based representation. In contrast, GPT-4o-mini achieved the lowest overall classification performance among the investigated models, particularly in terms of false positive predictions and reduced overall classification stability compared with the other LLMs. Furthermore, the obtained confusion matrices indicate that the proposed framework successfully distinguished between benign and malicious SDN traffic behaviors with relatively limited misclassification rates across most investigated models. The ROC curve analysis additionally demonstrates strong discrimination capability under the natural language-based traffic representation, where most investigated LLMs achieved AUC values close to 1.00. Overall, the obtained results indicate that the natural language-based traffic representation remains highly effective for retrieval-guided SDN traffic analysis and Carpet-Bombing DDoS attack detection within the proposed RAG-based framework.

#### 4.6.3 Overall Comparative Discussion

The overall comparative evaluation results demonstrate that both traffic representation strategies enabled highly effective Carpet-Bombing DDoS attack detection within the proposed RAG-based framework across all investigated LLMs. However, the structured JSON-based traffic representation generally achieved more stable and consistent classification performance compared with the natural language-based representation. This behavior can be attributed to the structured organization of traffic features within the JSON format, which reduces semantic ambiguity and provides more consistent contextual information during retrieval and inference operations. As a result, most investigated LLMs achieved lower false positive and false negative rates under the structured JSON representation.

In contrast, although the natural language-based traffic representation also achieved excellent overall detection performance, certain investigated LLMs exhibited slightly higher variations in classification behavior and inference stability. Since the natural language-based representation relies on descriptive textual traffic statements, differences in semantic interpretation across LLM architectures may influence retrieval consistency and contextual traffic understanding during inference operations. Nevertheless, the obtained results indicate that the proposed RAG-based framework remains highly effective under both traffic representation strategies.

Furthermore, the comparative evaluation demonstrates that the proposed framework configuration utilizing the Gemma-4-31B-IT model consistently achieved the strongest overall detection performance across both traffic representation strategies. In addition, the obtained results confirm that integrating structured traffic representation, similarity-based retrieval, and LLM-driven contextual traffic analysis provides a highly effective approach for intelligent Carpet-Bombing DDoS attack detection in SDN environments.

### 4.7 Comparative Analysis with Existing Studies

Table 6 presents a qualitative comparison between the proposed framework and representative studies related to DDoS detection and mitigation in SDN environments. The comparison focuses on several important design aspects, including SDN deployment environments, Carpet-Bombing DDoS attack analysis, LLM integration, retrieval-augmented analysis, interface-level traffic monitoring, training requirements, and real-time mitigation capabilities. This comparison is intended to highlight the architectural and operational characteristics of the proposed framework rather than providing a direct numerical performance comparison, since existing studies were evaluated under different datasets, experimental settings, and network environments.

Table 6: Qualitative comparison between the proposed framework and existing DDoS detection studies in SDN environments.

| Study | SDN-Based Environment | Carpet-Bombing Focus | LLM-Based Detection | Retrieval-Augmented Analysis | Interface-Level Features | Fine-Tuning / Training Required | Real-Time Mitigation |
|---|---|---|---|---|---|---|---|
| [19] | No | Yes | Yes | No | No | Yes | No |
| [20] | No | Yes | Yes | No | No | No | No |
| [21] | No | Yes | No | No | No | Yes | No |
| [22] | Yes | No | Yes | No | Yes | No | Yes |
| [23] | Yes | No | Yes | No | No | Yes | No |
| [24] | Yes | No | Yes | No | No | Yes | No |
| [25] | Yes | No | No | No | No | Yes | Yes |
| [26] | Yes | No | No | No | No | Yes | Yes |
| [27] | Yes | No | No | No | No | Yes | Yes |
| [28] | Yes | No | No | No | No | Yes | Yes |
| [29] | Yes | No | No | No | No | Yes | Yes |
| [30] | No | No | Yes | No | No | Yes | No |
| [31] | Yes | No | No | No | No | Yes | No |
| [32] | Yes | No | No | No | No | Yes | No |
| [33] | No | No | No | No | No | Yes | No |
| [34] | Yes | No | No | No | No | Yes | No |
| This Study | Yes | Yes | Yes | Yes | Yes | No | Yes |

As illustrated in Table 6, a large portion of existing studies primarily focus on conventional DDoS attack detection in SDN environments using traditional machine learning, deep learning, or fine-tuned LLM-based approaches. Although several studies have demonstrated high detection accuracy, most existing methods depend on supervised training procedures or fine-tuning operations that require labeled datasets, repeated retraining, and computationally expensive model adaptation processes. In contrast, the proposed framework utilizes a retrieval-augmented analysis mechanism combined with LLM-driven contextual inference to perform intelligent traffic classification without relying on conventional supervised model training or fine-tuning procedures.

In addition, only a limited number of existing studies specifically investigate Carpet-Bombing DDoS attacks, while most prior works primarily focus on conventional flooding attacks or general malicious traffic detection scenarios. Furthermore, existing Carpet-Bombing DDoS attack studies are commonly evaluated using general network traffic environments rather than SDN-specific infrastructures. Unlike previous approaches, the proposed framework is specifically designed for real-time Carpet-Bombing DDoS attack detection and mitigation within SDN environments using interface-level traffic analysis and contextual semantic retrieval mechanisms.

Moreover, most existing approaches rely primarily on flow-level traffic analysis and focus mainly on attack detection without providing integrated real-time mitigation capabilities. In contrast, the proposed framework combines retrieval-driven contextual traffic analysis, interface-level monitoring, and real-time mitigation mechanisms to rapidly detect and mitigate distributed Carpet-Bombing DDoS attacks before they significantly impact SDN network operation. Overall, the comparison demonstrates that the proposed framework provides a more adaptive and intelligent detection architecture by integrating retrieval-augmented analysis, LLM-based reasoning, and real-time mitigation capabilities within SDN environments.

## 5. Conclusion and Future Work

This paper presented a RAG-based framework for detecting and mitigating Carpet-Bombing DDoS attacks in SDN environments using semantic traffic retrieval and LLM-based inference. The proposed framework integrates contextual traffic retrieval, embedding generation, FAISS similarity search, and LLM to perform intelligent traffic classification without relying on conventional supervised training approaches. The experimental results demonstrated that the proposed framework achieved highly accurate detection performance under both structured JSON-based and natural language-based traffic representation strategies. In addition, comparative evaluation using multiple LLMs showed that the framework configuration utilizing the Gemma-4-31B-IT model achieved the strongest overall detection performance. Furthermore, the real-time experiments confirmed the capability of the proposed framework to rapidly detect and mitigate Carpet-Bombing DDoS attacks under different attack intensities while maintaining stable SDN network operation.

For future work, the proposed framework can be applied to larger SDN environments and evaluated under more diverse real-world network conditions and attack scenarios. In addition, future studies may investigate the integration of the proposed RAG-based framework with distributed multi-controller SDN architectures to further analyze its scalability and operational behavior in large-scale networks.

**Funding:** This work was supported by the National Natural Science Foundation of China under Grant No. 62171291, the Shenzhen Key Fundamental Research Funding under Grant No. JCYJ20220818100810023, and the Digital Asset Innovation Research Center Shenzhen University.